\documentclass[10pt]{wlscirep}
\usepackage[utf8]{inputenc}
\usepackage[T1]{fontenc}
\usepackage{newtxtext,newtxmath}
\usepackage[symbol]{footmisc}
\usepackage[linesnumbered,ruled,vlined]{algorithm2e}
\usepackage{graphicx}
\usepackage{bm}
\usepackage{amsmath}
\usepackage{multirow}
\usepackage{listings}
\usepackage{caption}
\usepackage{subcaption}
\usepackage[figuresleft]{rotating}
\title{Tracking an Untracked Space Debris After an Inelastic Collision Using Physics Informed Neural Network}

\author[1,*]{Harsha M.}
\author[2]{Gurpreet Singh}
\author[3]{Vinod Kumar}
\author[1]{Arun Balaji Buduru}
\author[1]{Sanat K. Biswas}
\affil[1]{Indraprastha Institute of Information Technology Delhi, New Delhi, 110020, India}
\affil[2]{U R Rao Satellite Centre, ISRO, Bengaluru, 560071, India}
\affil[3]{Indian National Space Promotion and Authorization Center, Ahmedabad, 380058, India}
\affil[*]{harsham@iiitd.ac.in}

\begin{abstract}
With the sustained rise in satellite deployment in Low Earth Orbits, the collision risk from untracked space debris is also increasing. Often small-sized space debris (below 10 cm) are hard to track using the existing state-of-the-art methods. However, knowing such space debris' trajectory is crucial to avoid future collisions. We present a Physics Informed Neural Network (PINN) - based approach for estimation of the trajectory of space debris after a collision event between active satellite and space debris. In this work, we have simulated 8565 inelastic collision events between active satellites and space debris. To obtain the states of the active satellite, we use the TLE data of 1647 Starlink and 66 LEMUR satellites obtained from \href{space-track.org}{space-track.org}. The velocity of space debris is initialized using our proposed velocity sampling method, and the coefficient of restitution is sampled from our proposed Gaussian mixture-based probability density function. Using the velocities of the colliding objects before the collision, we calculate the post-collision velocities and record the observations. The state (position and velocity), coefficient of restitution, and mass estimation of un-tracked space debris after an inelastic collision event along with the tracked active satellite can be posed as an optimization problem by observing the deviation of the active satellite from the trajectory. We have applied the classical optimization method, the Lagrange multiplier approach, for solving the above optimization problem and observed that its state estimation is not satisfactory as the system is under-determined. Subsequently, we have designed Deep Neural network-based methods and Physics Informed Neural Network (PINN) based methods for solving the above optimization problem. We have compared the performance of the models using root mean square error (RMSE) and interquartile range of the predictions. It has been observed that the PINN-based methods provide a better estimation performance for position, velocity, mass and coefficient of restitution of the space debris compared to other methods.

\end{abstract}

\begin{document}

\flushbottom
\maketitle
%
%
\thispagestyle{empty}

\section*{Introduction}

Since the launch of Sputnik in 1957, approximately 13,000 satellites have been deployed in the Low Earth orbit \cite{montaruliAdaptiveTrackEstimation2022}. With collisions, explosions and fragmentations, the number of space debris sized less than 1 cm is now estimated to be more than 1,000,000\cite{montaruliAdaptiveTrackEstimation2022}. The US Space Surveillance Network has a catalogue of about 30,000 resident space objects only\cite{montaruliAdaptiveTrackEstimation2022}. Since 97\% of space debris is not tracked, they pose a greater threat to active satellites as well as future space missions.

With this large number of space debris, the risk of in-orbit collision has increased, and the number of collision events has seen a rise in the past decade. Collision in space can be destructive, for example, Iridium-Cosmos collision\cite{tanAnalysisIridium332013} of 2009, or non-destructive, for example, Canadarm2 of the International Space Station hit by a small piece of space debris in 2021 \cite{dattaOpedDamageCanadarm22021}. Another incident which was catalogued as a non-destructive collision was in 2013 when Blits satellite was possibly hit by un-tracked space debris \cite{kelsoWhatHappenedBLITS2013}. In non-destructive collision events, fragmentation does not happen. While there has been exhaustive research on modelling destructive collisions \cite{braun2020drama,braun2021recent}, on risk assessment due to small space debris \cite{lopez-calleComparisonCubesatMicrosat2023}, and reconnecting fragments in space to their parent satellite body\cite{cellettiReconnectingGroupsSpace2021}, the extraction of trajectory information of untracked space debris observing a non-destructive collision has not been studied extensively. 

Recently, Harsha et al.\cite{mDeepNeuralNetworkbased2023} formulated a space debris position, velocity and mass estimation problem considering the position and velocity deviation of an active satellite as observation with non-destructive and elastic collision assumption. The performance of the classical estimation methods as well as deep learning (DL) based approaches, were examined for this problem. It was observed that the position, velocity, and mass estimation performance of the Ensemble Neural Network-based technique are similar to those of classical methods. It should be noted that the elastic collision assumption in the above preliminary study was idealistic. The above estimation problem becomes complex when a more rigorous inelastic collision model is considered where the momentum is conserved, but the kinetic energy is not conserved. Hence, it is of interest to examine the trajectory estimation performance of both classical and ML-based methods under the inelastic collision assumption. 


Deep Neural Network (DNN)-based techniques have already been proposed for asteroid exploration, spacecraft rendezvous and terrain navigation \cite{song_deep_2022}. The convolutional neural network (CNN) has been proposed for pose estimation of uncooperative spacecraft \cite{becktor2022robust} and for object detection and tracking in space \cite{petit2012vision}. ESA's Hera mission is planning to carry out image processing of double asteroid system Dimorphos-Didymos using CNN\cite{kaluthantrigeCNNbasedImageProcessing2023} and demonstrate optical navigation. This mission will also investigate the impact of kinetic impactor released by  NASA's Double asteroid redirection test (DART) mission using deep learning techniques\cite{stevensonPredictingEffectsKinetic2022}. 


Deep learning techniques have also been demonstrated to carry out initial screening of conjunction assessment among 170 million space-object pairs \cite{stevensonBenchmarkingDeepLearning2023}, which otherwise is a computation intensive process in the domain of Space Situational Awareness. Automatic collision avoidance maneuver have been proposed using ML algorithms\cite{sanchezIntelligentDecisionSupport2023} combining with Dempster-Shafer theory of evidence.
Deep Neural network achieve satisfactory performance when trained with large amounts of data. However, these standard neural network-based methods inevitably face the challenge of drawing conclusions and making decisions under partial information, where the data acquisition is costly and scarce while dealing with complex and non-linear physical systems\cite{Raissi_2018}.
In addition, the solution of the traditional neural network does not guarantee adherence to the underlying physical properties in problems involving physical systems. To address these issues, the Physics informed Neural network (PINN) was proposed \cite{Raissi_2018}. PINN \cite{raissiPhysicsinformedNeuralNetworks2019,karniadakis2021physics} uses the laws governing the system dynamics in the loss function as prior information. Note that the training of a neural network essentially implies finding the set of weights and biases for all the nodes in the network, which minimises the loss function. Hence, the inclusion of the system dynamics based on physical laws in the loss function act as a constraint for the output while training the neural network using the training data set. The PINNs can be used as surrogates in learning models used for autonomous navigation, uncertainty quantification and any real-time application that need inference \cite{goswami2022physicsinformed}.

In this article, we have studied the problem of tracking unknown space debris, observing an inelastic and non-destructive collision event. Additionally, we assumed the space debris was in a stable orbit before the collision. Under the inelastic collision assumption, the coefficient of restitution \cite{brakeComprehensiveSetImpact2017a} is an unknown parameter in addition to the variables that need to be estimated under the elastic collision assumption. We have examined the performance of the classical estimation technique and various DNN and PINN-based methods for estimating the position, velocity, and mass of the space debris and the coefficient of restitution for the inelastic collision for the above-mentioned problem using 8235 collision simulations for training and 330 collision simulations for testing. The inelastic collisions were simulated using the model described by Schwager et al.\cite{schwagerCoefficientRestitutionLinear2007} The performance of the DNN and PINN-based methods are compared with the classical method for the inelastic and non-destructive collision. Towards tracking unknown space debris, we have made the following key contributions:
\begin{enumerate}
    \item Formulation of space debris tracking problem observing a non-destructive and inelastic collision: We have posed the unknown space debris tracking problem after an inelastic event as a position, velocity, mass, and the coefficient of restitution estimation problem considering the active satellite position and velocity deviation as observations.
    \item Formulation of an appropriate physics loss function and application of PINN: We have formulated a physics loss function for the above problem and trained a PINN using the developed loss function to solve the above estimation problem.
    \item A velocity sampling method for simulating random in-orbit collision events: For the training of DNN and PINN-based models as well for testing the performance of the ML-based approaches and the classical approach, we simulated a total of 8565 inelastic collision events. Note that, for a collision event, the active satellite position and the space debris position will be approximately the same at the time of the collision, while the velocities will be different. However, any random velocity vector at the given position does not necessarily result in a stable elliptical orbit. We have proposed a velocity sampling method to generate the velocity vector of space debris for a given active satellite position at the time of the collision, which guarantees an elliptical orbit with a periapsis above a given minimum allowable altitude from the mean sea level.
\end{enumerate}


The rest of the article is organised as follows: We have formally defined the problem in the next section, and then discussed the classical estimation method and various DNN-based approaches and provided the physics loss function for the PINN for solving the problem. Then we have discussed the collision data generation method, including the formulation of the velocity sampling algorithm and simulation of collision in space. Next, we have compared the performance of the classical estimation technique as well as the gradient boosting, DNN and PINN-based methods for solving the estimation problem. We conclude the article by consolidating our observations and delineating the future research direction.

\section*{Problem Definition}
\label{problemDefinition}
\begin{figure}
    \centering
    \includegraphics[width=0.6\textwidth]{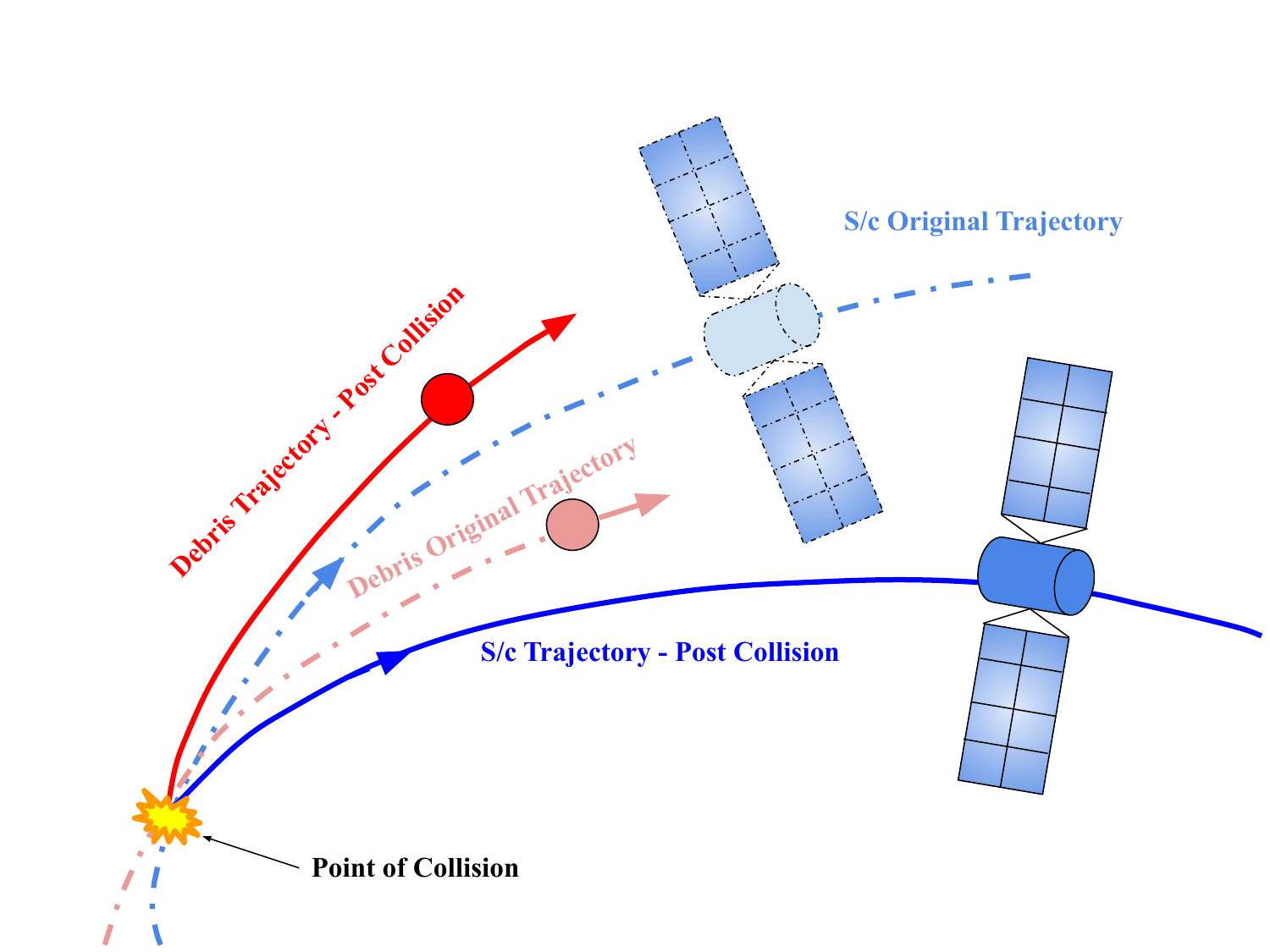}
    \caption{Illustration of a collision in space}
    \label{fig:collision}
\end{figure}

Consider a satellite and space debris undergoing an inelastic collision in space as depicted in Fig. \ref{fig:collision}.
Our objective is to find the trajectory of the space debris by observing the position and velocity deviation of the active satellite after the collision.
Let $\bm{r}_{sat}$ and  $\bm{v}_{sat}$ be the position and velocity of the satellite, and $\bm{r}_{d}$ and $\bm{v}_{d}$ be the position and velocity of the space debris. We assume satellite and space debris are spherical for ease of the analysis. Let $t_c^-$ and $t_c^+ $ denote the time before collision and time after collision respectively. At the time of the collision, the position of the satellite and the position of debris can be written as
\begin{equation}
    \bm{r}_{sat}(t_c^+) = \bm{r}_{sat}(t_c^-) 
\end{equation}
\begin{equation}
    \bm{r}_{d}(t_c^+) = \bm{r}_{d}(t_c^-) = \bm{r}_{sat}(t_c^-) - (\rho_{sat} + \rho_{d}) \bm{\hat{v}}_{rel}
    \label{eq:rso_pos_after}
\end{equation}
where $\rho_{sat}$ and $\rho_{d}$ are radii of the satellite and debris, respectively and $\bm{\hat{v}}_{rel}$ is the unit vector along the relative velocity of debris with respect to the satellite and is given by
\begin{equation}
    \bm{\hat{v}}_{rel} = \frac {\bm{v}_{sat}(t_c^-) - \bm{v}_{d}(t_c^-) }  {|| \bm{v}_{sat}(t_c^-) - \bm{v}_{d}(t_c^-) || }
\end{equation}
The velocity of the satellite and debris after an inelastic collision is \cite{brakeComprehensiveSetImpact2017a}
\begin{equation}
    \bm{v}_{sat}(t_c^+) = \frac {m_{sat}\bm{v}_{sat}(t_c^-) + m_{d}\bm{v}_{d}(t_c^-) +  \epsilon m_{d} (\bm{v}_{d}(t_c^-) - \bm{v}_{sat}(t_c^-))}{m_{sat} + m_{d}}
    \label{eq:sat_vel_after}
\end{equation}
\begin{equation}
    \bm{v}_{d}(t_c^+) = \frac {m_{d}\bm{v}_{d}(t_c^-) + m_{sat}\bm{v}_{sat}(t_c^-) +  \epsilon m_{sat} (\bm{v}_{sat}(t_c^-) - \bm{v}_{d}(t_c^-))}{m_{sat} + m_{d}}
    \label{eq:rso_vel_after}
\end{equation}
where $\bm{v}_{sat}(t_c^-) $ indicates velocity of satellite just before the collision and $\bm{v}_{sat}(t_c^+)$ indicates velocity of satellite just after the collision and $\epsilon$ is the coefficient of restitution \cite{schwagerCoefficientRestitutionLinear2007} of the inelastic collision. Then change in velocity of the active satellite can be written as 
\begin{equation}
    \Delta\bm{v}_{sat}(t_c^+) = \bm{v}_{sat}(t_c^+) - \bm{v}_{sat}(t_c^-)
\end{equation}
Define the measurement $\bm Z$ as
\begin{equation} \label{eq:measZ}
    \bm Z = \begin{bmatrix} \bm{r}_{sat}(t_c^+) \\ \Delta\bm{v}_{sat}(t_c^+) \end{bmatrix}  + \omega(t_c^+) 
     =  \begin{bmatrix} \bm{r}_{sat}(t_c^+) \\ \bm{v}_{sat}(t_c^+) - \bm{v}_{sat}(t_c^-) \end{bmatrix}  + \omega(t_c^+) 
\end{equation}
where $\omega(t_c^+)$ is noise in measurement. As the active satellite is being tracked, $ \bm{v}_{sat}(t_c^+)$ and $ \bm{v}_{sat}(t_c^-)$ can be measured, and therefore measurement $\bm{Z}$ is available. Using equations \eqref{eq:rso_pos_after}, \eqref{eq:sat_vel_after} and \eqref{eq:rso_vel_after}, $\bm{Z}$ can be written as 
\begin{equation}
    \bm Z = f(\bm{r}_{d}(t_c^-), \bm{v}_{d}(t_c^-), m_{d}, \epsilon) + \omega(t_c^+)
    \label{eq:func_f}
\end{equation}
Obtaining $\bm{r}_{d}(t_c^-), \bm{v}_{d}(t_c^-), m_{d}$ and $\epsilon$ from $\bm Z$ is essentially an estimation problem. Using the estimated parameters,  $\bm{v}_d(t_c^+)$ can be obtained and thus the trajectory of the space debris after the collision can be computed.

\section*{Classical approach}
\label{sec:classical_approach}
One can estimate $\bm{r}_{d}(t_c^-), \bm{v}_{d}(t_c^-), m_{d}$ and $\epsilon$ from observation $\bm Z$ using the Least Squared Estimation (LSE):
\begin{equation}
 \widehat{\bm{r}}_{d}(t_c^-), \widehat{\bm{v}}_{d}(t_c^-), \widehat{m}_{d}, \widehat{\epsilon}= \underset{{\bm{r}_{d}(t_c^-), \bm{v}_{d}(t_c^-), m_{d}, \epsilon}}{\arg\min} (\bm{\Delta Z}^T\bm{\Delta Z})
\end{equation}
where $\widehat{\bm{r}}_{d}(t_c^-), \widehat{\bm{v}}_{d}(t_c^-), \widehat{m}_{d}, \widehat{\epsilon}$ are the estimate of the desired quantities and
\begin{equation}
    \bm{\Delta Z} = \bm{Z} - f(\widehat{\bm{r}}_{d}(t_c^-), \widehat{\bm{v}}_{d}(t_c^-), \widehat{m}_{d}, \widehat{\epsilon}) 
\end{equation}
Note that the above estimation problem formulation does not consider any mass constraints for the debris. However, we are particularly interested in the untracked debris, which is of small size and essentially has a very small mass. The inclusion of the mass constraint transforms the original problem into a constrained optimization problem, which can be solved using the Lagrange Multiplier approach \cite{mDeepNeuralNetworkbased2023}. The Lagrangian $\mathcal{L}$ for this constrained optimization problem can be defined as
\begin{equation}
    \mathcal{L} = \bm{\Delta Z}^T \bm{\Delta Z} + \lambda_1(1 - m_{d}) + \lambda_2(m_{d} - \delta m)\label{eq:lagrangian}
\end{equation}
which includes the mass constraint $\delta m<m_{d}<1~kg$. Here, $\lambda_1$ and $\lambda_2$ are Lagrange multipliers. The position, velocity, mass and coefficient of restitution of the debris can be estimated by solving
\begin{align}
    \begin{bmatrix}\frac{\partial \mathcal{L}}{\partial \bm{r}_{d}} & \frac{\partial \mathcal{L}}{\partial \bm{v}_{d}} & \frac{\partial \mathcal{L}}{\partial m_{d}} & \frac{\partial \mathcal{L}}{\partial \epsilon}\end{bmatrix} &= \bm{0}\\
    \begin{bmatrix}\frac{\partial \mathcal{L}}{\partial \lambda_1} & \frac{\partial \mathcal{L}}{\partial \lambda_2}\end{bmatrix} &= \bm{0}
    \label{eq:lagrange_method}
\end{align}
There are various techniques available to solve this minimisation problem. We have used the gradient descent approach to solve the problem numerically.

\section*{Deep Learning-based approaches}
\label{sec:ML_approaches}
\begin{figure}[ht]
    \centering
    \includegraphics[width=0.4\textwidth]{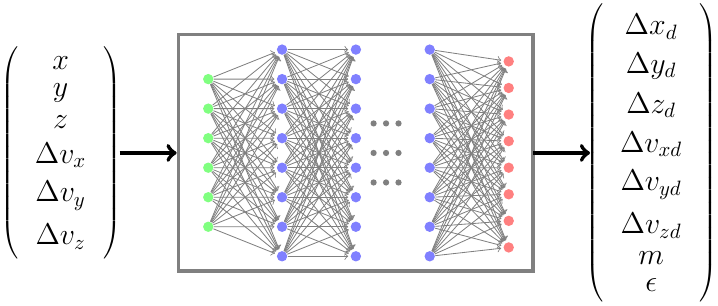}
    \caption{Deep Neural Network (DNN) architecture}
    \label{fig:DNN}
\end{figure}

\begin{figure}[ht]
    \centering
    \includegraphics[width=0.6\textwidth]{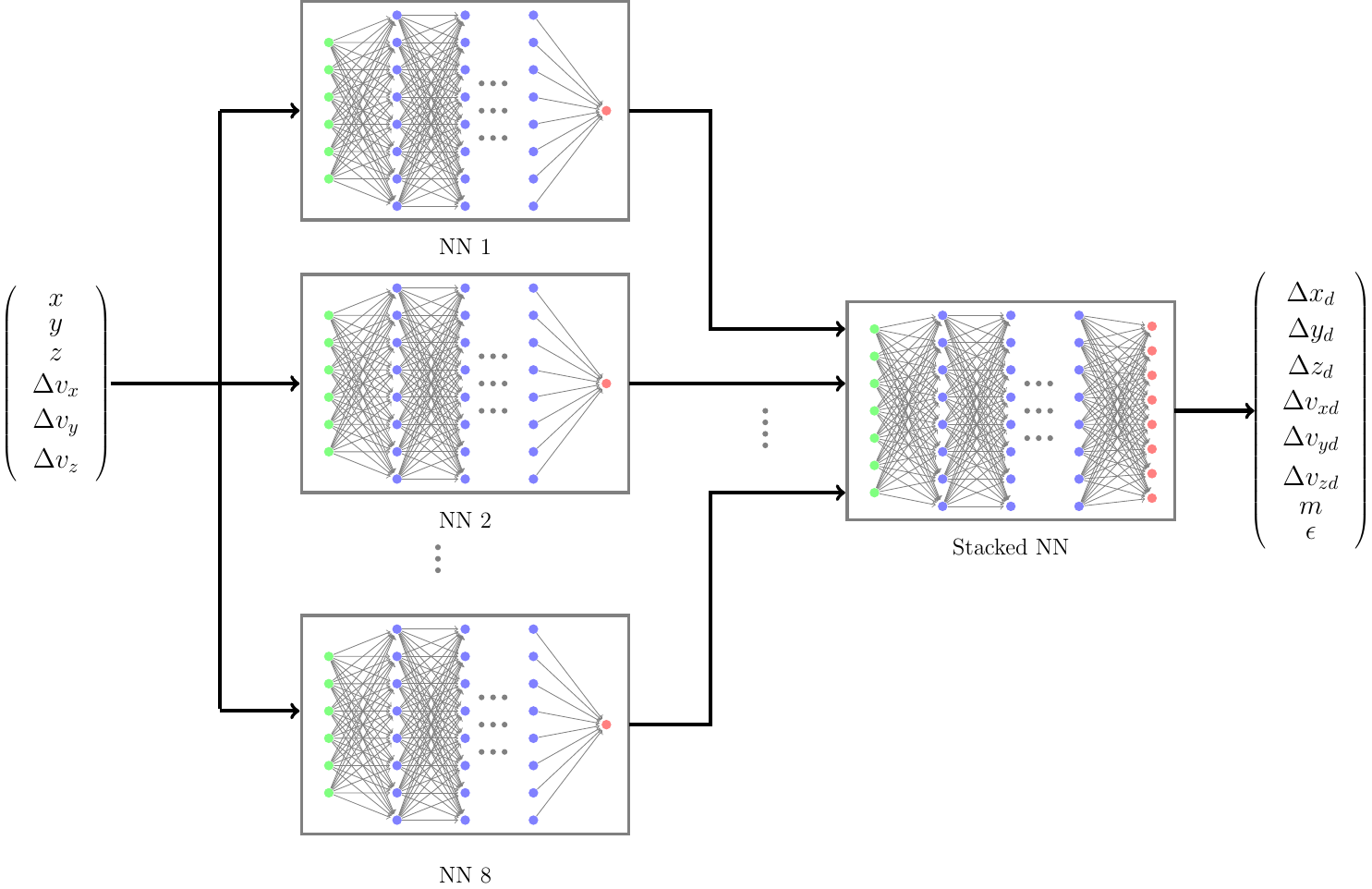}
    \caption{Stacked Deep Neural Network (StackDNN) architecture}
    \label{fig:EnsDNN}
\end{figure}

An alternate approach for estimating the position, velocity, mass and coefficient of restitution of the debris after a collision can be based on a DNN model. Using \eqref{eq:func_f} one can write
\begin{equation}
    E[\bm{r}_{d}(t_c^-), \bm{v}_{d}(t_c^-), m_{d}, \epsilon] = f^{-1}(\bm{Z})
    \label{eq:inverse_f}
\end{equation}
considering $\bm{\omega}(t_c^+)$ as a zero mean noise vector. Essentially the trained DNN model approximates the inverse function $f^{-1}(\cdot)$. The DNN model can be trained using simulated collision data. We consider $\bm{Z}$ as the input to the DNN and $\bm{\Delta r}, \bm{\Delta v}, m_{d}, \epsilon$ as the output of the DNN, where
\begin{equation}
    \bm{\Delta r} = \begin{bmatrix}\Delta x_d \\ \Delta y_d\\ \Delta z_d\end{bmatrix} = \bm{r}_{d}(t_c^-) - \bm{r}_{sat}(t_c^-)
\end{equation}
and
\begin{equation}
    \bm{\Delta v} = \begin{bmatrix}\Delta v_{xd} \\ \Delta v_{yd}\\ \Delta v_{zd}\end{bmatrix} = \bm{v}_{d}(t_c^-) - \bm{v}_{sat}(t_c^-)
\end{equation}
We consider two distinct DNN architectures for this problem. The first architecture is the usual DNN with an input layer with 6 inputs, multiple hidden layers and an output layer with 8 outputs as shown in Fig. \ref{fig:DNN}. The second architecture comprises of 8 parallel DNNs with 6 inputs, multiple hidden layers and 1 output. Essentially, each of these parallel DNNs learns one of the 8 outputs. The outputs of these 8 parallel DNNs are connected to another DNN with 8 inputs and 8 outputs. The last DNN block ensures learning the dependency of each output variable with other output variables. The architecture is shown in Fig. \ref{fig:EnsDNN}.

We refer to this architecture as the stacked Deep Neural Network (StackDNN). For both architectures, Mean Squared Error (MSE) loss function is used for training the models. We define the MSE loss for this problem as
\begin{align}
    &\mathcal{L}_{MSE}(\widehat{\bm{r}}_{d}, \widehat{\bm{v}}_{d}, \hat{m}_{d}, \hat{\epsilon})\nonumber\\
    &=\frac{1}{N}\sum_{i=1}^N\left[(\bm{r}_{d_i} - \widehat{\bm{r}}_{d_i})^T(\bm{r}_{d_i} - \widehat{\bm{r}}_{d_i}) + (\bm{v}_{d_i} - \widehat{\bm{v}}_{d_i})^T(\bm{v}_{d_i} - \widehat{\bm{v}}_{d_i}) + (m_{d_i} - \hat{m}_{d_i})^2 + (\epsilon_i - \hat{\epsilon_i})^2\right]
\end{align}
where $(\cdot)_{d_i}$ and $\widehat{(\cdot)}_{d_i}$ denote the output data for the $i^{th}$ collision event and the corresponding predicted output data using the DNN. $N$ is the number of collision events used to generate the training data.

\section*{Physics Informed Neural Network}
\label{sec:PINN}
Having discussed the DNN-based method for estimating the position, velocity, mass and coefficient of restitution of debris after the collision, it is essential to note that the estimated position and velocity must be the solution to the differential equation that governs the motion of the debris. Based on Newton's Law of Gravitation, the debris dynamics can be written as
\begin{equation}
    \begin{bmatrix}\dot{\bm r}_{d}\\\dot{\bm v}_{d}\end{bmatrix} = \begin{bmatrix}\bm v_{d}\\ -\frac{\mu}{r_{d}^3}\bm{r}_{d}\end{bmatrix}
    \label{eq:sat_dynamics}
\end{equation}
However, DNN-based solutions do not guarantee the above constraint. The Physics Informed Neural Network\cite{raissiPhysicsinformedNeuralNetworks2019} preserves the laws of physics in the solution by incorporating a physics loss in the loss function used to train the DNN. Using \eqref{eq:sat_dynamics}, we defined the physics loss for the PINN as
\begin{align}
    &\mathcal{L}_{phy}(\widehat{\bm{r}}_{d}, \widehat{\bm{v}}_{d})\nonumber\\
    &= \frac{1}{N}\sum_{i=1}^N\left[\left(\frac{\mu}{r_{d_i}^3}\bm{r}_{d_i} - \frac{\mu}{\hat{r}_{d_i}^3}\widehat{\bm{r}}_{d_i}\right)^T\left(\frac{\mu}{r_{d_i}^3}\bm{r}_{d_i} - \frac{\mu}{\hat{r}_{d_i}^3}\widehat{\bm{r}}_{d_i}\right)
    + \left(\bm{v}_{d_i} - \widehat{\bm{v}}_{d_i}\right)^T\left(\bm{v}_{d_i} - \widehat{\bm{v}}_{d_i}\right)\right]
    \label{eq:physics_loss}
\end{align}
for the problem under consideration. The significance of this loss function is that the minimisation of $\mathcal{L}_{phy}(\widehat{\bm{r}}_{d}, \widehat{\bm{v}}_{d})$ minimises the difference between the acceleration computed using the output of the training data and the PINN output, and the difference between the velocity computed using the output of the training data and the PINN output. As a result, this loss function minimises the distance between the derivative of the vector $\left[\bm{r}_{d}~~\bm{v}_{d}\right]^T$ computed using the training data, and the derivative of the same computed using the PINN predicted output. Since the velocity loss is already included in $\mathcal{L}_{phy}(\widehat{\bm{r}}_{d}, \widehat{\bm{v}}_{d})$ we define the MSE loss for the PINN as
\begin{equation}
    \mathcal{L}_{MSEP}(\widehat{\bm{r}}_{d}, \hat{m}_{d}, \hat{\epsilon})
    =\frac{1}{N}\sum_{i=1}^N\left[(\bm{r}_{d_i} - \widehat{\bm{r}}_{d_i})^T(\bm{r}_{d_i} - \widehat{\bm{r}}_{d_i}) + (m_{d_i} - \hat{m}_{d_i})^2 + (\epsilon_i - \hat{\epsilon_i})^2\right]
    \label{eq:msep_loss}
\end{equation}
The complete loss function for the PINN is
\begin{equation}
    \mathcal{L}_{PINN} = \mathcal{L}_{phy}(\widehat{\bm{r}}_{d}, \widehat{\bm{v}}_{d}) + \mathcal{L}_{MSEP}(\widehat{\bm{r}}_{d}, \hat{m}_{d}, \hat{\epsilon})
\end{equation}
We use this loss function in both DNN and StackDNN for performance comparison.

\section*{Collision data generation}
Having discussed the DNN, StackDNN, PINN, and StackPINN approaches for tracking the debris after an inelastic collision event, let us now turn to the preparation of the training data. Publicly available data on inelastic and non-destructive collision events are very few and not sufficient to train the neural networks. Hence simulated collision data are required for training the deep learning models. 

The position and velocity information of active satellites can be obtained from publicly available catalogues. However, the position and velocity of the debris which collides with an active satellite at a given epoch must be simulated. The simulated debris must be in a complete orbit, i.e. the orbit eccentricity must be less than 1. Additionally, to form a stable orbit, the periapsis of the debris cannot be at a very low altitude. 

The debris position is given by \eqref{eq:rso_pos_after} at the time of the collision. From \eqref{eq:rso_pos_after}, we can approximate that the debris position as nearly equal to the active satellite position because $r_{sat}\gg \rho_{sat}, \rho_{d}$. Corresponding to this position, infinitely many velocity vectors are possible for which the debris will be on a complete orbit. However, the magnitude and the direction of the velocity vector can not be any arbitrary magnitude and direction. In the next subsection, we describe a method of sampling the velocity of debris for a given position vector in space and a given minimum periapsis altitude. 
\subsection*{Velocity sampling}
\label{sec:vel_sample}
\begin{figure}[ht]
    \begin{minipage}[b]{0.48\linewidth}
        \centering
        \includegraphics[width = 0.9\textwidth]{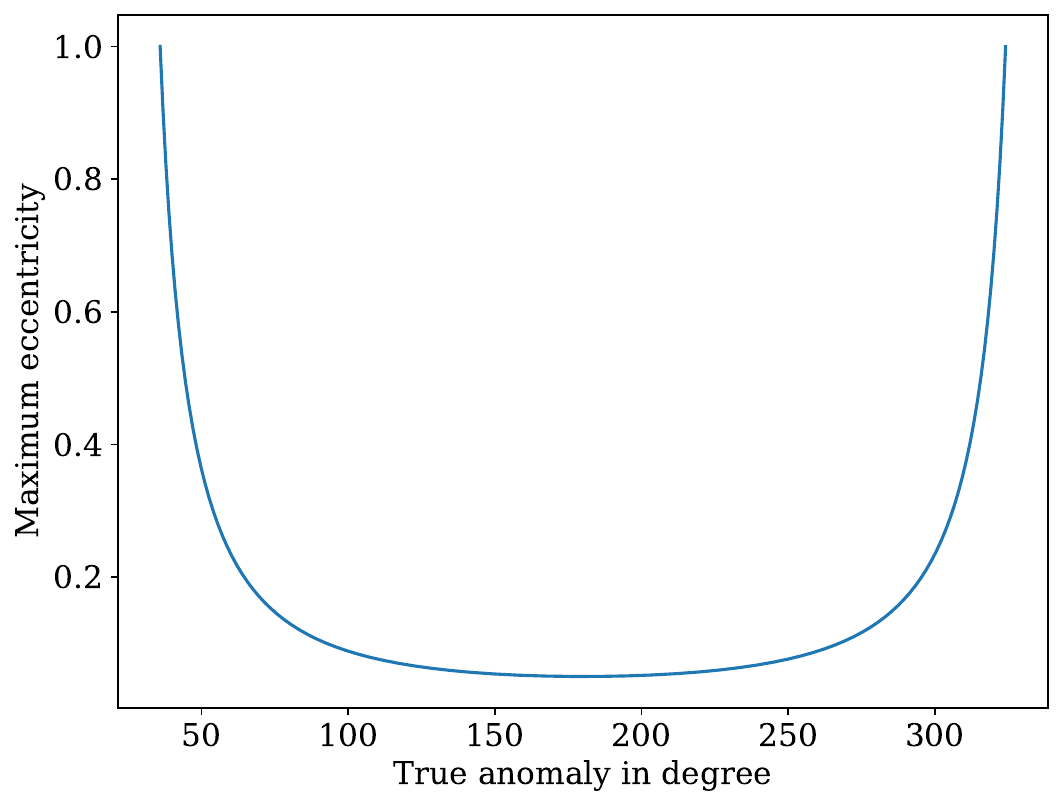}
        \caption{Maximum eccentricity vs. True anomaly}
        \label{fig:ecc_vs_ta}
    \end{minipage}
    \hfill
    \begin{minipage}[b]{0.5\linewidth}
        \centering
        \includegraphics[width=0.7\textwidth]{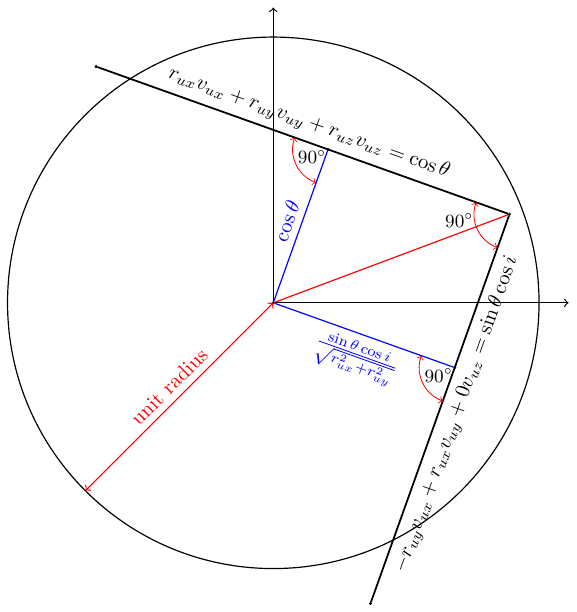}
        \caption{Crossection of the intersection of the unit sphere and planes \eqref{eq:cos_theta} and \eqref{eq:sin_theta}}
        \label{fig:crossectional_geometry}
    \end{minipage}
\end{figure}

Consider $r$ is the norm of the active satellite as well as the debris position vector $\bm{r}$ at the time of the collision, $h$ is the norm of the angular momentum vector $\boldsymbol{h}$ of the colliding debris, $e$ is the eccentricity of the debris orbit,  $\phi$ is the true anomaly of the debris during the collision, $R_e$ is the radius of the Earth and $a_{min}$ minimum allowable periapsis altitude. If the semi-major axis for the debris orbit is $a$, then the periapsis $r_p = a(1 - e)$. Also
\begin{equation}
    \frac{h^2}{\mu} = r_p(1 + e)
\end{equation}
and
\begin{equation}
    r = \frac{h^2}{\mu}\frac{1}{1 + e\cos{\phi}}\label{eq:keplar}
\end{equation}
Considering $R_{min} = R_e + a_{min}$ as the minimum allowable distance from the earth centre
\begin{equation}
    r_p = \frac{r(1 + e\cos{\phi})}{1 + e}>R_{min}\label{eq:periapsis}
\end{equation}
then
\begin{equation}
   0<e< \frac{r - R_{min}}{R_{min} - r\cos{\phi}}<1 \label{eq:ecc_lim}
\end{equation}
and
\begin{equation}
    \cos^{-1}\frac{2R_{min}-r}{r}<\phi < 2\pi - \cos^{-1}\frac{2R_{min}-r}{r}\label{eq:ta_lim}
\end{equation}
Fig. \ref{fig:ecc_vs_ta} shows the variation of maximum possible eccentricity with the true anomaly.
Now, the position of the debris at the time of collision is
\begin{equation}
    \bm{r}_{d} = \bm{r}= r\begin{bmatrix}r_{ux}\\r_{uy}\\r_{uz}\end{bmatrix}\label{eq:pos_unit_vec}
\end{equation}
and velocity of the debris is 
\begin{equation}
    \bm{v}_{d} = v\begin{bmatrix}v_{ux}\\v_{uy}\\v_{uz}\end{bmatrix}\label{eq:vel_unit_vec}
\end{equation}
where $\begin{bmatrix}r_{ux}&r_{uy}&r_{uz}\end{bmatrix}^T$ and $\begin{bmatrix}v_{ux}&v_{uy}&v_{uz}\end{bmatrix}^T$ are the position and velocity unit vectors respectively, and
\begin{equation}
    v = \sqrt{\frac{\mu}{r}\left(2 - \frac{1-e^2}{1+e\cos{\phi}}\right)}\label{eq:v_mag}
\end{equation}
The angle between $\bm{r}$ and $\bm{v}$ is $\theta$ \cite{vallado2001fundamentals}
\begin{equation}
    \theta = \sin^{-1}\frac{1 + e\cos{\phi}}{\sqrt{1+2e\cos{\phi}+ e^2}}\label{rv_angle}
\end{equation}
The angular momentum vector of the debris is
\begin{align}
    \bm{h} &= \bm{r}\times\bm{v}\nonumber\\
 &= rv\begin{bmatrix}r_{uy}v_{uz} - r_{uz}v_{uy}\\ r_{uz}v_{ux} - r_{ux}v_{uz}\\r_{ux}v_{uy} - r_{uy}v_{ux}\end{bmatrix}\label{eq:angular_m_vector}
\end{align}
and
\begin{equation}
    h = rv\sin{\theta}\label{eq:angular_m_mag}
\end{equation}
then the cosine of the inclination $i$ of the orbit
\begin{equation}
    \cos{i} = \frac{rv(r_{ux}v_{uy} - r_{uy}v_{ux})}{rv\sin{\theta}}\label{eq:inclination}
\end{equation}
Consequently, the velocity unit vector needs to satisfy the following:
\begin{gather}
    r_{ux}v_{ux} + r_{uy}v_{uy} +r_{uz}v_{uz} = \cos{\theta}\label{eq:cos_theta}\\
    - r_{uy}v_{ux} + r_{ux}v_{uy} + 0v_{uz} = \sin{\theta}\cos{i}\label{eq:sin_theta}\\
    \sqrt{v_{ux}^2 + v_{uy}^2 + v_{uz}^2} = 1\label{eq:unit_sphere}
\end{gather}

Geometrically, the condition for forming a complete orbit is - the line of intersection of planes \eqref{eq:cos_theta} and \eqref{eq:sin_theta} must lie within the unit sphere defined by \eqref{eq:unit_sphere}. We can write \eqref{eq:sin_theta} in the normal form:
\begin{equation}
   - \frac{r_{uy}}{\sqrt{r_{ux}^2+r_{uy}^2}}v_{ux} + \frac{r_{ux}}{\sqrt{r_{ux}^2+r_{uy}^2}}v_{uy} + 0\cdot v_{uz} = \frac{\sin{\theta}\cos{i}}{\sqrt{r_{ux}^2+r_{uy}^2}}\label{eq:sin_theta_normal} 
\end{equation}

Since $\sqrt{r_{ux}^2 + r_{uy}^2 + r_{uz}^2} = 1$, \eqref{eq:cos_theta} is already in the normal form. Observing \eqref{eq:cos_theta} and \eqref{eq:sin_theta_normal}, one can deduce that the angle between the two planes is $90^\circ$. Then from the cross-sectional geometry shown in Fig. \ref{fig:crossectional_geometry}, the criteria for the intersection of the two planes within the unit sphere is
\begin{equation}
    \cos^2{\theta} + \frac{\sin^2{\theta}\cos^2{i}}{r_{ux}^2 + r_{uy}^2}\leq1\label{eq:geometric_condition}
\end{equation} 
Then
\begin{equation}
  -\sqrt{\frac{r_{ux}^2 + r_{uy}^2}{\sin^2{\theta}(1-\cos^2{\theta})}}\leq\cos{i}\leq\sqrt{\frac{r_{ux}^2 + r_{uy}^2}{\sin^2{\theta}(1-\cos^2{\theta})}}\label{eq:inc_lim}  
\end{equation}
Using the ranges of $\phi$, $e$ and $\cos{i}$ given by \eqref{eq:ta_lim}, \eqref{eq:ecc_lim} and \eqref{eq:inc_lim} one can independently sample these quantities and subsequently solve for $v_{ux}$, $v_{uy}$ and $v_{uz}$ from \eqref{eq:cos_theta}, \eqref{eq:sin_theta} and \eqref{eq:unit_sphere}. Note that the solution will result in two unit vectors for velocity. Finally, the velocity magnitude can be calculated using \eqref{eq:v_mag}. Algorithm \ref{alg:v_Sample} describes the velocity sampling steps.
\begin{algorithm}[ht]
    \SetAlgoLined
    \caption{Velocity sampling}
    \label{alg:v_Sample}
    \SetKwInOut{Input}{Input }
    \SetKwInOut{Output}{Output }
    \Input{$\bm{r}$, minimum altitude, Number of samples}
    \Output{Velocity samples}
    \For{i=1:Number of samples}
    {
        Given the allowed minimum altitude, find the minimum and maximum true anomaly $\phi_{min}$ and $\phi_{max}$ for the given position vector using \eqref{eq:ta_lim}\;
        Sample $\phi \sim U(\phi_{min}, \phi_{max})$\;
        For a sampled $\phi$ calculate maximum eccentricity $e_{max}$ using \eqref{eq:ecc_lim}\;
        Sample $e \sim U(0, e_{max})$\;
        Calculate $\theta$\;
        Compute the lower bound $\cos{i}_{min}$ and upper bound $\cos{i}_{max}$ for $\cos{i}$ using \eqref{eq:inc_lim}\;
        Sample $\cos{i} \sim U(\cos{i}_{min}, \cos{i}_{max})$\;
        Solve for $v_{ux}, v_{uy}, v_{uz}$ using \eqref{eq:cos_theta}, \eqref{eq:sin_theta} and \eqref{eq:unit_sphere}\;
        Calculate $v$ using \eqref{eq:v_mag}
    }
\end{algorithm}

\begin{figure}[ht]
    \begin{minipage}[b]{0.48\linewidth}
    \centering
    \includegraphics[width = 0.9\textwidth]{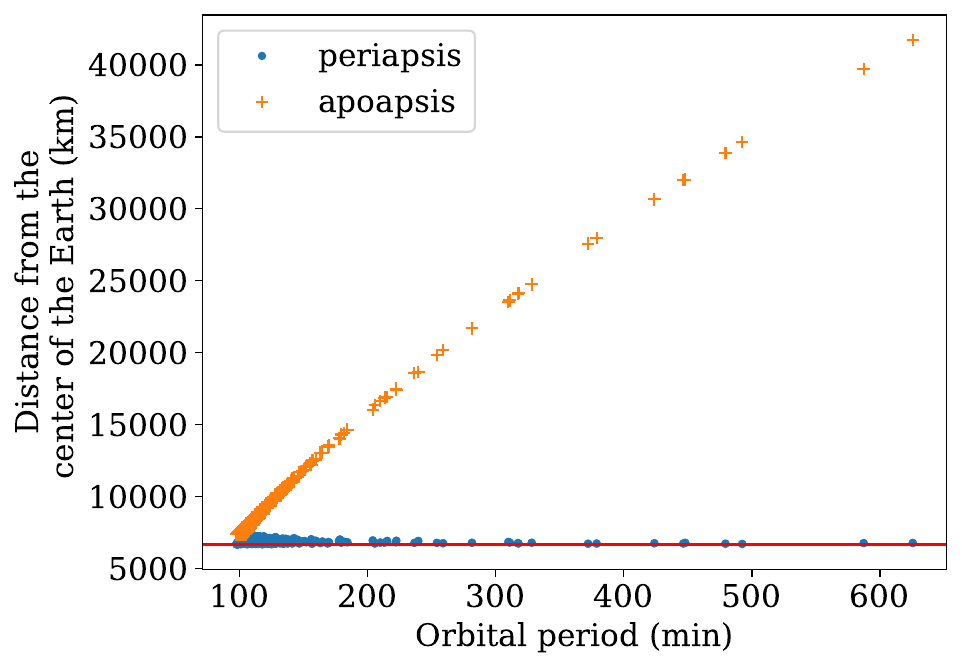}
    \caption{Gabbard chart for the generated velocity samples}
    \label{fig:gabbard_Samples}
    \end{minipage}
    \begin{minipage}[b]{0.5\linewidth}
    \centering
    \includegraphics[width = 0.6\textwidth]{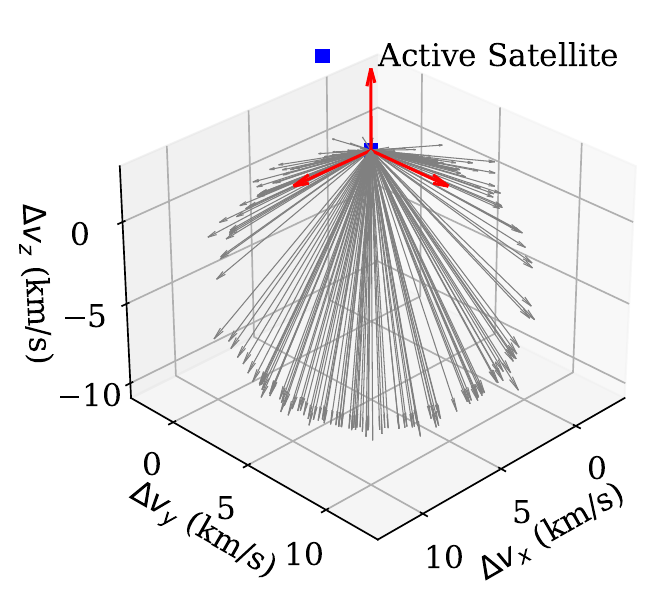}
    \caption{Relative velocities for the sampled space debris velocity with respect to the active satellite}
    \label{fig:velocity_distr}
    \end{minipage}
\end{figure}
Using this algorithm, we have sampled 2000 velocity vectors for an active satellite position at
\begin{equation}
    \bm{r} = \begin{bmatrix}3664.75536654& 3893.60049258& 5062.87536598\end{bmatrix}^T km
\end{equation}
and a minimum perigee altitude of 300 km. Fig. \ref{fig:gabbard_Samples} shows that the generated samples have perigees at altitudes higher than 300 km. Fig. \ref{fig:velocity_distr} shows the relative velocities for the samples with respect to the active satellite. It can be deduced from Fig. \ref{fig:velocity_distr} that collision cannot occur from any arbitrary direction for a given position vector in space.

\subsection*{Simulation of the collision}
\label{sec:simulation}
We have generated inelastic collision observations considering the collision model described in the Problem Formulation section. We have used Python 3.7 with astropy\cite{theastropycollaborationAstropyProjectSustaining2022} and poliastro\cite{rodriguezPoliastroAstrodynamicsLibrary2016} libraries for orbit simulations of the active satellites as well as the debris. 

We have considered 5 collision epochs $t_c$ each at 1200 UTC, starting from April 29, 2023, to May 3, 2023. We consider 1647 satellite orbits from the Starlink constellation and 66 satellite orbits from the LEMUR constellation for generating collision simulations. The Two Line Element (TLE) data for each constellation were obtained from \href{www.spacetrack.org}{www.spacetrack.org} on April 27, 2023. Using TLE data, we have calculated the time difference $\Delta t$ between the collision time $t_c$ and the the TLE time, and propagated the active satellites. The orbits are propagated considering J2 and J3 zonal harmonics and the exponential atmospheric drag model.

\begin{figure}[ht]
    \centering
    \includegraphics[width=\textwidth]{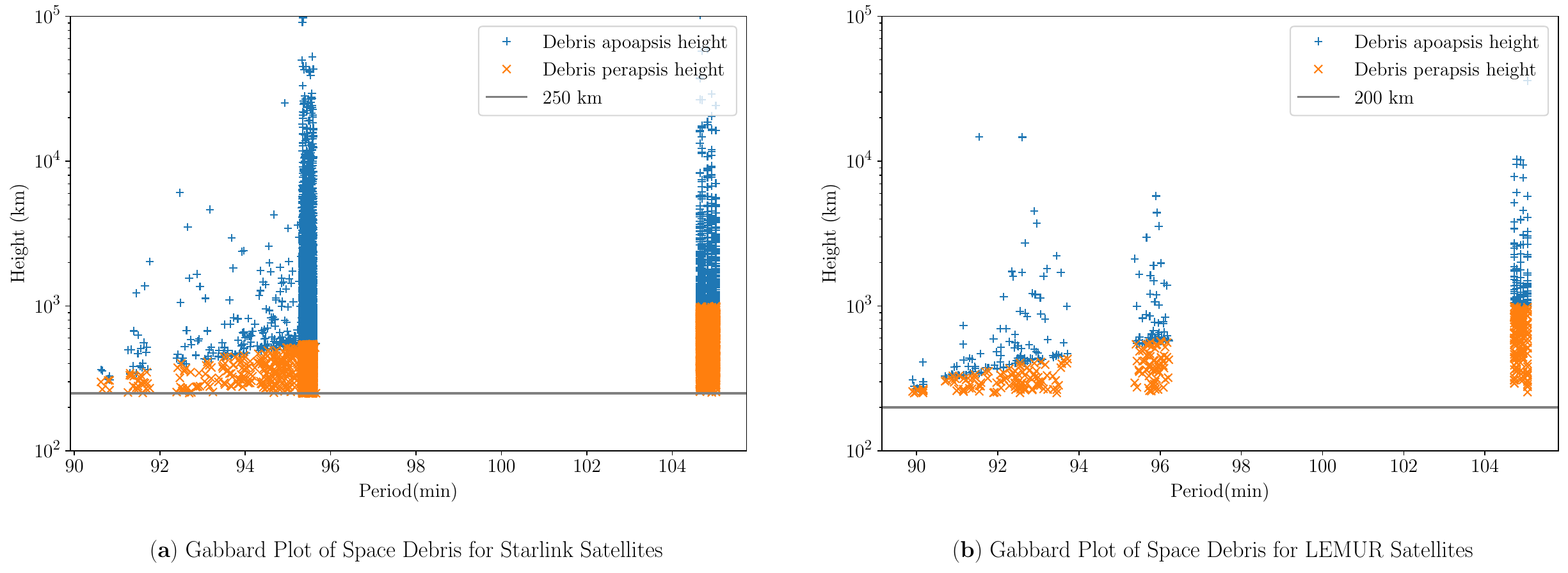}
    \caption{Gabbard plot of space debris}
    \label{fig:gabbard_rso}
\end{figure}

For the debris simulations, we sampled the mass of each untracked debris from a normal distribution with 0.5 kg mean and varying standard deviations of (0 kg, 0.001 kg, 0.002 kg, 0.003 kg, 0.004 kg, 0.005 kg, 0.006 kg). For simplicity, satellites and debris are considered spherical objects of radius 0.5 m and 0.1 m, respectively, and the mass of the satellite is 100 times that of sampled mass of debris. The position of the debris \bm{$r_{d}$} is considered to be the same as \bm{$r_{sat}$}, whereas the velocity of the debris $\bm{v}_{d}$ is obtained using algorithm \ref{alg:v_Sample}. The minimum perigee altitude for the debris corresponding to the Starlink satellites was set at 250 km, and for LEMUR, it was 200 km. The Gabbard plots for the debris generated for the Starlink and LEMUR constellations are shown in Fig. \ref{fig:gabbard_rso}.


\begin{figure}[ht]
    \centering
    \includegraphics[width=0.6\textwidth]{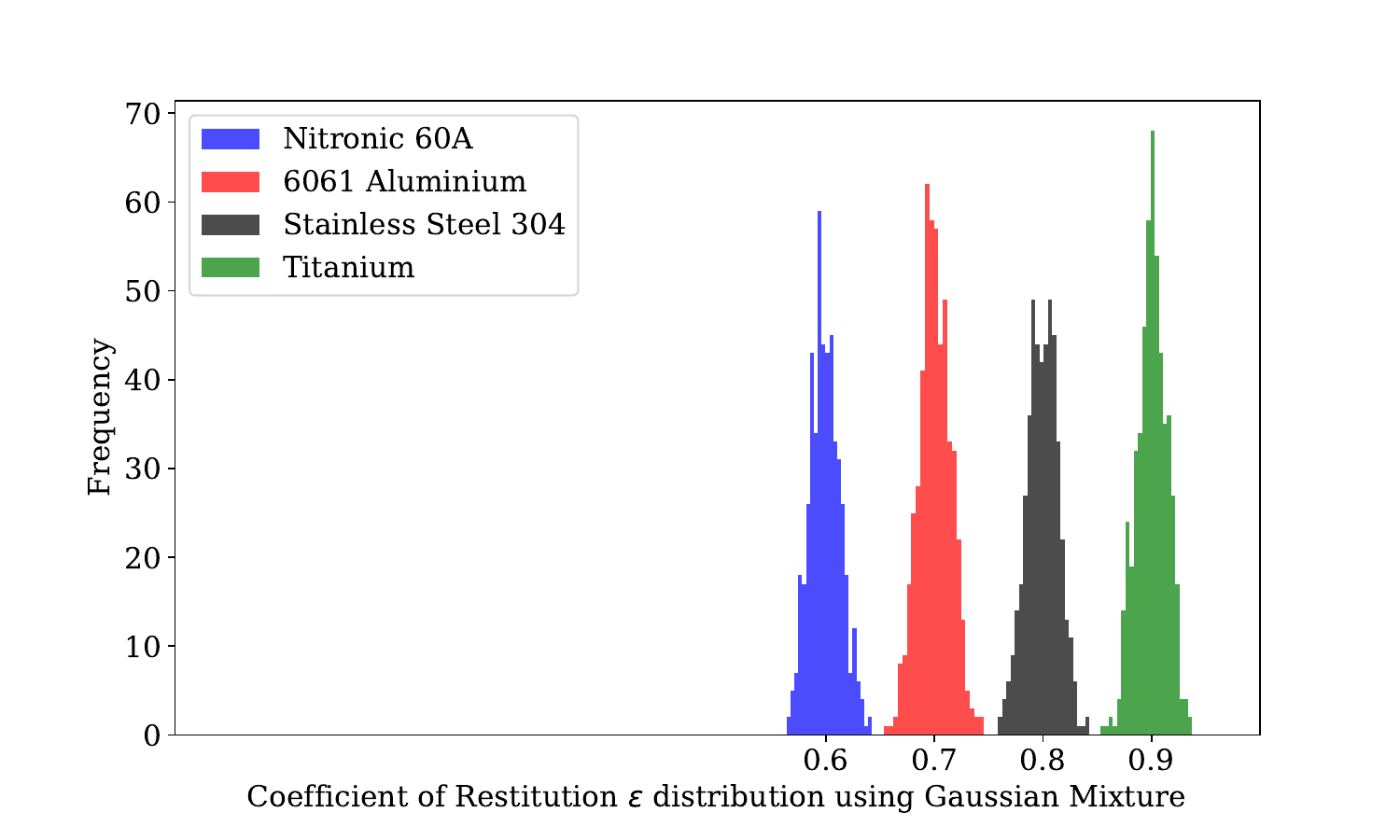}
    \caption{Coefficient of restitution ($\epsilon$) distribution using Gaussian mixture}
    \label{fig:CoR_3sigma}
\end{figure}

Coefficient of restitution ($\epsilon$) values for each of the collisions have been sampled from a Gaussian mixture distribution, as shown in Fig. \ref{fig:CoR_3sigma}. It consists of 4 Gaussian distributions with 0.015 standard deviations sampled around uniformly distributed means ranging from 0.5 to 1 where the mean has been assumed based on the $\epsilon$ values of the satellite materials (6061-T6 Aluminium, Stainless Steel 304, Nitronic 60A and Titanium) \cite{brakeComprehensiveSetImpact2017a}. 

Inelastic collision simulations are generated using the position, velocity, mass and coefficient of restitution values for the respective satellite and debris pairs in equations \eqref{eq:sat_vel_after} and \eqref{eq:rso_vel_after}. Post-collision at time $t_c^+$, the observations are generated using equation \eqref{eq:measZ}. Here, the noise $\omega(t_c^+)$ signifies the tracking uncertainty of the active satellite. We consider varying standard deviation (0m, 100 m, 200 m, 300 m, 400 m, 500 m) for position and (0 m/s, 50 m/s, 100 m/s, 150 m/s, 200 m/s, 250 m/s) for velocity in each axis.

A total of 8235 collision events were simulated for the orbits of the Starlink constellation using Param Pravega Super Computer. 66 collision events with 6 varying noise factors were simulated for the LEMUR constellation using a local workstation. For each collision, we have recorded $\bm{Z}$ and $\bm{r}_{d}$, $\bm{v}_{d}$, $m_{d}$ and $\epsilon$. We consider the collision data corresponding to the orbits of the Starlink constellation as the training data and the same corresponding to the orbits of the LEMUR constellation as the test data. 

\section*{Results and discussion}\label{sec:results}
We used the training data described in the previous section for training the DNN, PINN, StackDNN, and StackPINN models. We trained four different models with 2, 4, 6, and 8 hidden layers for each of the ML models, with 5 nodes in each hidden layer. Rectified Linear Unit (ReLU) is selected as the activation function for each node. We have used adaptive moment (adam) optimizer with the learning rate equal to 1e-3 for all the models. All the models are trained for 50 epochs with a batch size equal to 32. We have also trained the Extreme Gradient Boosting (XGBoost)\cite{chen2016xgboost} regressor model with the same training data.

\begin{figure}
    \centering
    \includegraphics[width=\textwidth]{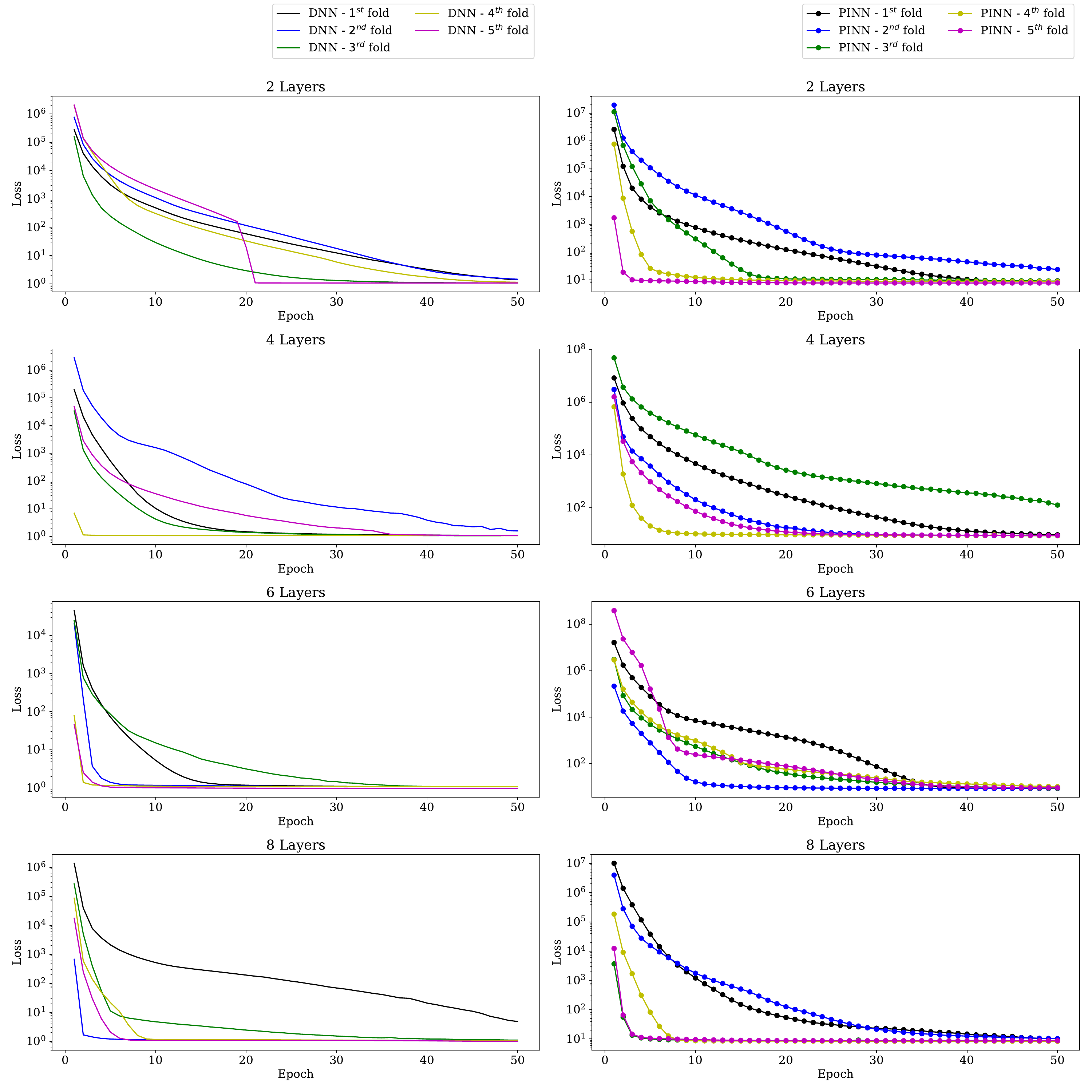}
    \caption{Training loss for DNN and PINN models in 5-Fold Cross-validation using 8235 samples and 50 epochs}
    \label{fig:dnn_pinn_loss_plot(fulldata)}
\end{figure}

\begin{figure}
    \centering
    \includegraphics[width=\textwidth]{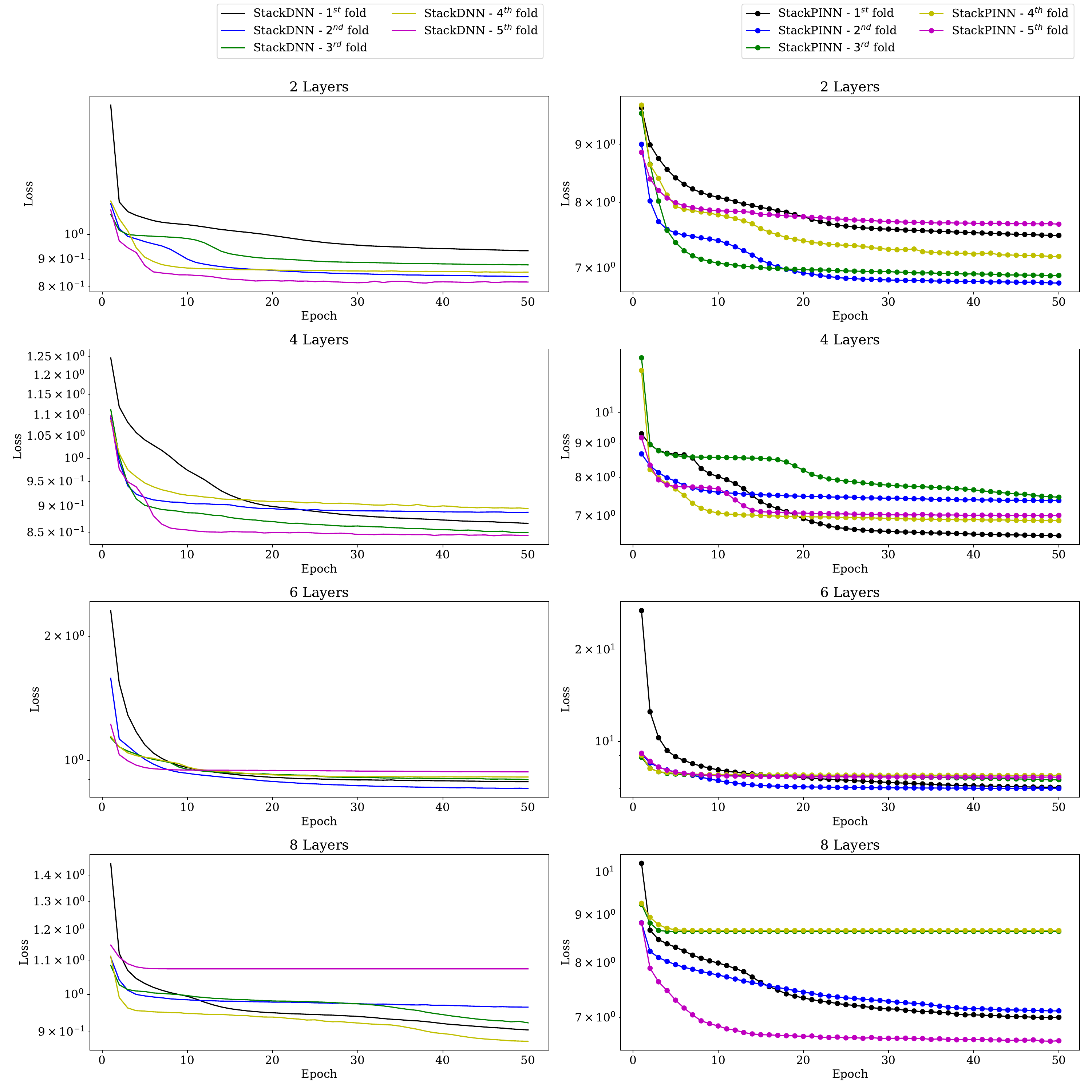}
    \caption{Training loss for StackDNN and StackPINN models in 5-Fold Cross-validation using 8235 samples and 50 epochs}
    \label{fig:stackdnn_stackpinn_loss_plot(fulldata)}
\end{figure}

The training is performed in a workstation with 8 Intel Xeon processor cores and 62GB of RAM. The training of all the ML models is done using 5-fold cross-validation for different sizes of training data. All the ML models are trained with 8235 samples and 4000 samples{\footnote[2]{\label{note}All results analysed during the current study are available in the \textbf{collisionAI} repository, \href{https://github.com/HarshaSSL/collisionAI}{https://github.com/HarshaSSL/collisionAI}}}. Fig. \ref{fig:dnn_pinn_loss_plot(fulldata)} shows the change in training loss with respect to training epochs for the DNN and the PINN models trained using 8235 samples. Similarly, Fig. \ref{fig:stackdnn_stackpinn_loss_plot(fulldata)} shows the change in training loss with respect to training epochs for the StackDNN and the StackPINN models trained using 8235 samples. It is observed in most of the folds, for models with an increased number of hidden layers, the training epochs decreases for the training loss to converge. In StackDNN and StackPINN architectures (Fig. \ref{fig:EnsDNN}), the input to stacked NN is obtained from already trained parallel models. This reduces the initial loss value to the order of $10^1$ when compared to DNN and PINN models’ initial loss value which is of the order $10^7$.The results for models trained using 4000 samples can be found in the supplementary information.

After analysing the performance of the trained models, we pick the best-performing model based on the lowest RMSE for the validation data during the 5-fold cross-validation. Inference is done using the selected model for the prediction of position, velocity, mass, and coefficient of restitution of inelastic collisions of the test dataset generated using LEMUR satellites. 


\begin{figure}
    \centering
    \includegraphics[width=\textwidth]{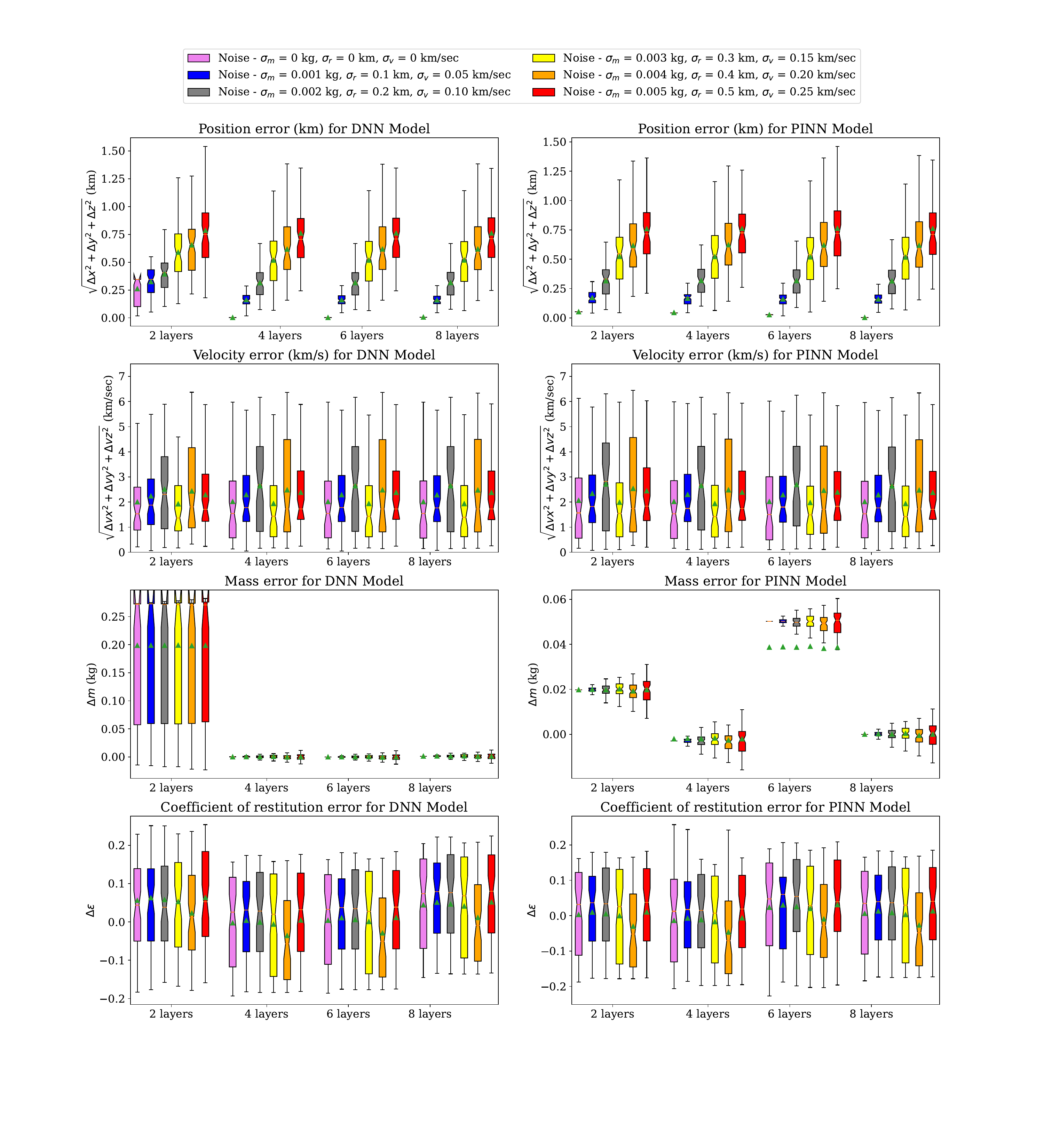}
    \caption{Error comparison of DNN and PINN models trained using 8235 samples and 50 epochs for multiple noise variations in LEMUR Satellites collision data}
    \label{fig:DNN(full)_error_vs_noise}
\end{figure}
\begin{figure}
    \centering
    \includegraphics[width=\textwidth]{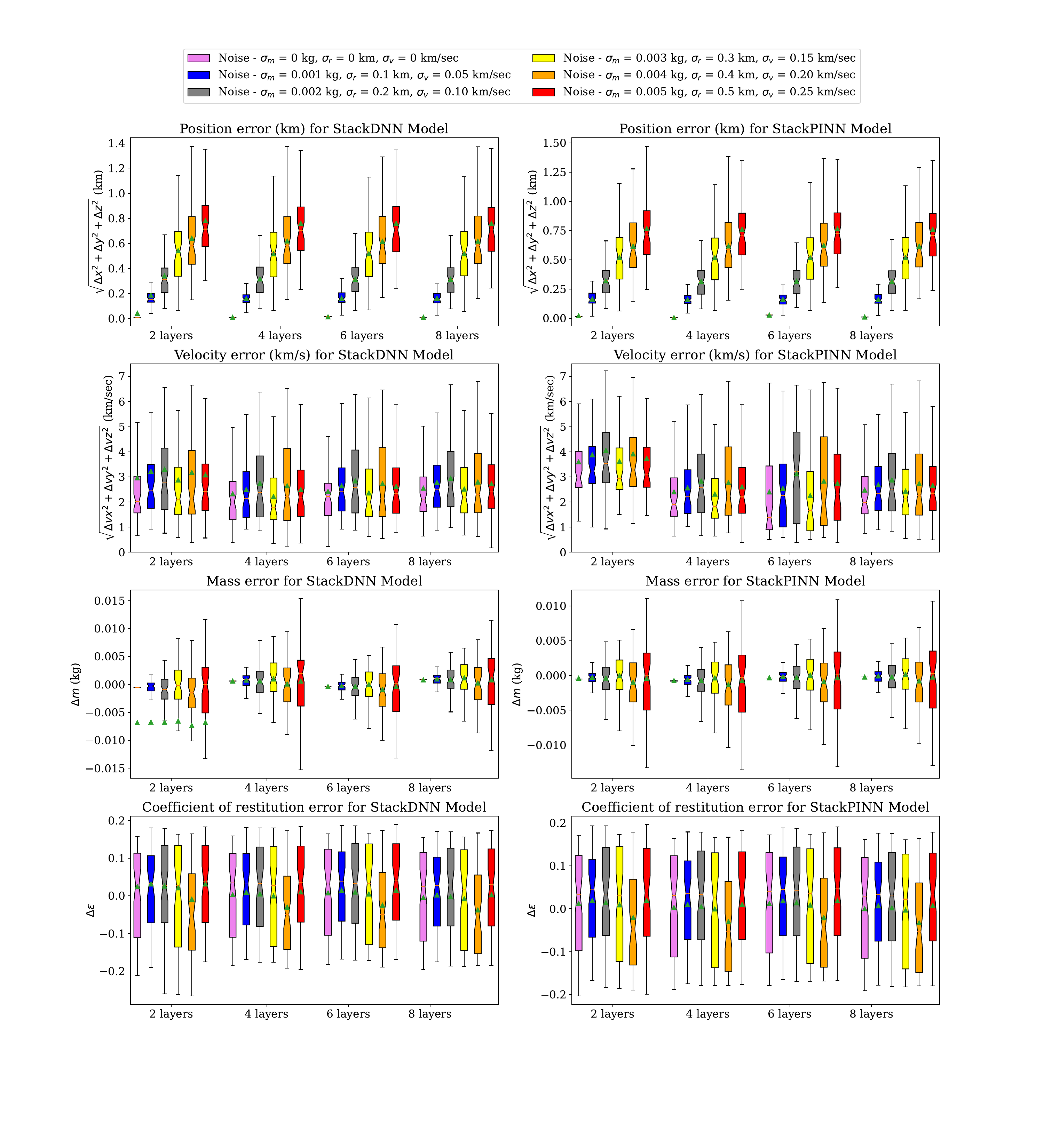}
    \caption{Error comparison of StackDNN and StackPINN models trained using 8235 samples and 50 epochs for multiple noise variations in LEMUR satellites inelastic collision data}
    \label{fig:StackDNN(full)_error_vs_noise}
\end{figure}
\begin{figure}
    \centering
    \includegraphics[width=0.8\textwidth]{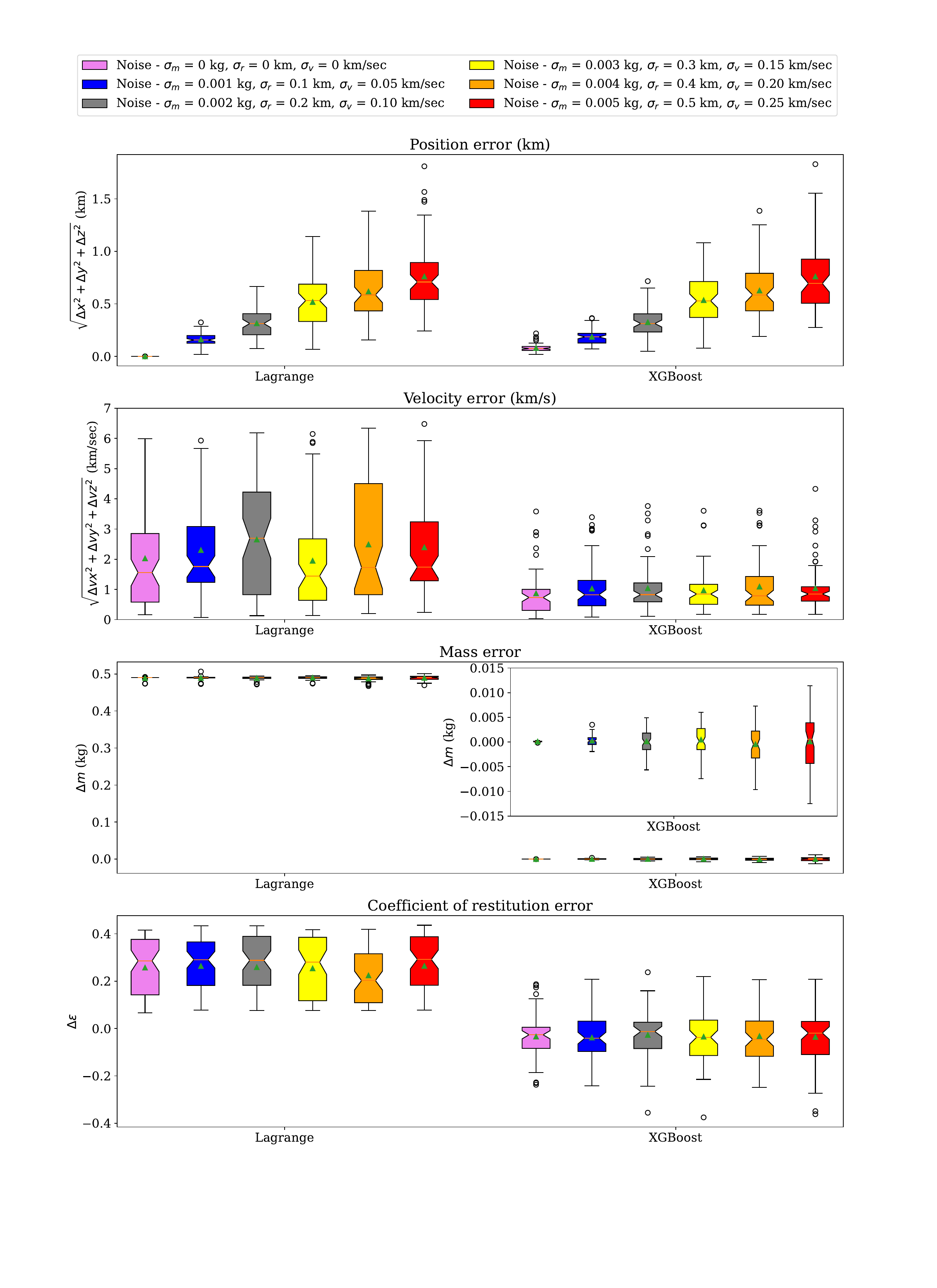}
    \caption{Error comparison of Lagrange method and XGBoost models trained using 8235 samples for multiple noise variations in LEMUR Satellites inelastic collision data}
    \label{fig:XGB_lagrange(full)_error_vs_noise}
\end{figure}

Using the observation $\bm{Z}$ for all 66 test collision events with 6 different noise standard deviations as mentioned in section \ref{sec:simulation} in the selected ML models, $\bm{r}_{d}(t_c^-)$, $\bm{v}_{d}(t_c^-)$, $m_{d}$ and $\epsilon$ were estimated. Additionally, the same variables were estimated using the Lagrange multiplier method for the same set of observations. We recorded the error in estimation for performance comparison. 

Fig. \ref{fig:DNN(full)_error_vs_noise} shows the box plots of the position, velocity, mass and coefficient of restitution estimation errors recorded using selected DNN and PINN models. The position and velocity estimation errors for all the methods are represented by the Euclidian norm of the x, y, and z axes in the ECI frame. It can be observed that for position estimation error, the median and the interquartile range increase for the test cases with increasing noise standard deviation. A similar trend in the interquartile range can be found in the mass estimation error plots . In position, velocity and coefficient of restitution error estimation plots, it is observed that the performance of DNN and PINN models with 4, 6, and 8 hidden layers does not improve compared to 2 layers. However, in mass estimation error plots, DNN and PINN with 8 layers provide marginally better results as these models with more hidden layers are better function approximators \cite{mhaskar2016deep}.

Fig. \ref{fig:StackDNN(full)_error_vs_noise} shows the box plots of the position, velocity, mass, and coefficient of restitution estimation errors recorded using selected StackDNN and StackPINN models. The results follow a similar trend as the DNN and PINN models. However, StackDNN and StackPINN provides marginally better mass estimation using 2 layers. Fig. \ref{fig:XGB_lagrange(full)_error_vs_noise} shows the box plots of estimation errors recorded using Lagrange and XGBoost models. The results of the Lagrange method and XGBoost models follow a similar trend to that of DNN and PINN models for different noise standard deviations. However, the median and interquartile range for the estimation errors of XGBoost models is the least among all the models. The median and interquartile range of the estimation errors of the Lagrange method are the highest among all the methods.We have found through experiments that the output is sensitive to position noise standard deviation compared to the standard deviation for velocity and mass noises 

The summary of 5-Fold cross-validation of the models with 2 hidden layers is shown in Table \ref{tab:RMSE_5fold}. The table shows the mean $\pm$ standard deviation of the Root Mean Squared Error (RMSE) for the validation data of the 5-fold cross-validation analysis. We have observed that the mean RMSE across all the training is higher when trained with smaller datasets. It is also observed that the StackPINN produces lesser variation in the RMSE when the smaller dataset is used.

\begin{table}[ht]
\centering
\caption{RMSE tabular summary for 5-fold cross validation of training the models with different number of samples}
\label{tab:RMSE_5fold}
\begin{tabular}{|c|l|c|c|c|c|}
\hline
\textbf{\begin{tabular}[c]{@{}c@{}}Number of\\ Train samples\end{tabular}} & \multicolumn{1}{c|}{\textbf{\begin{tabular}[c]{@{}c@{}}Model\end{tabular}}} & \textbf{$\Delta r$ (km)} & \textbf{$\Delta v$ (km/s)} & \textbf{$\Delta m$ (kg)} & \textbf{$\Delta \epsilon$} \\ \hline
\multirow{5}{*}{\textbf{4000}}                                             & \textbf{DNN}                                                                             & 5.2704 $\pm$ 7.0132        & 6.821 $\pm$ 6.0505             & 1.544 $\pm$ 1.8549         & 2.1735 $\pm$ 1.9109          \\
                                                                           & \textbf{PINN}                                                                            & 9.2218 $\pm$ 11.667       & 19.410 $\pm$ 21.356          & 3.8594 $\pm$ 4.8815        & 3.7805 $\pm$ 2.0651          \\
                                                                           & \textbf{StackDNN}                                                                        & 5.2704 $\pm$ 7.0132        & 6.821 $\pm$ 6.0505             & 1.5440 $\pm$ 1.8549         & 2.1735 $\pm$ 1.9109          \\
                                                                           & \textbf{StackPINN}                                                                       & 0.3264 $\pm$ 0.3149        & 3.0294 $\pm$ 0.1066            & 0.0909 $\pm$ 0.1328        & 0.1467 $\pm$ 0.0386          \\
                                                                           & \textbf{XGBoost}                                                                         & 0.1832 $\pm$ 0.0020         & 1.0837 $\pm$ 0.0325            & 0.001 $\pm$ 9.47e-06       & 0.0917$\pm$0.0013            \\ \hline
\multirow{5}{*}{\textbf{8235}}                                             & \textbf{DNN}                                                                             & 1.2813 $\pm$ 1.2237        & 3.3195 $\pm$ 0.6529            & 0.421 $\pm$ 0.4746         & 0.8052 $\pm$ 0.6608          \\
                                                                           & \textbf{PINN}                                                                            & 0.9443 $\pm$ 0.7666        & 3.0712 $\pm$ 0.2684            & 0.3405 $\pm$ 0.2371        & 0.4976 $\pm$ 0.4599          \\
                                                                           & \textbf{StackDNN}                                                                        & 1.2813 $\pm$ 1.2237        & 3.3195 $\pm$ 0.6529            & 0.421 $\pm$ 0.4746         & 0.8052 $\pm$ 0.6608          \\
                                                                           & \textbf{StackPINN}                                                                       & 0.1752 $\pm$ 0.0098        & 4.3062 $\pm$ 1.8048            & 0.0252 $\pm$ 0.0538        & 0.1177 $\pm$ 0.0062          \\
                                                                           & \textbf{XGBoost}                                                                         & 0.1771 $\pm$ 0.0010         & 0.9979 $\pm$ 0.0187            & 0.001 $\pm$ 1.863e-05      & 0.0854 $\pm$ 0.0010             \\ \hline
\end{tabular}
\end{table}

From the above discussion, it can be discerned that, for 2 hidden layers, the PINN improves the space debris position, mass, and coefficient of restitution estimation performance significantly compared to the Lagrange multiplier-based and DNN-based methods. The performance of the DNN becomes similar to the PINN for higher numbers of hidden layers. However, the results show no significant advantage to increasing the number of layers for the problem under consideration. In addition, the StackDNN and the StackPINN performances are similar for the different numbers of hidden layers and comparable to the PINN model's performance with 2 hidden layers. However, the stacked architecture is 9 different Neural Networks, increasing the complexity of the implementation.

The training time required for different folds during 5-fold cross-validation, inference time for the selected models, and stored memory for reusability for each selected model with 2 hidden layers is shown in Table \ref{tab:train_time_storage}. XGBoost model requires 4000 KB of memory, which is 100 times more than the memory required to store PINN models.

\begin{table}[ht]

\centering
\caption{Tabular results of models consisting of time taken for training during 5-Fold cross validation, inference time for predicting and evaluating the LEMUR collision data (Nominal Noise) and their respective stored memory for reusability.}
\label{tab:train_time_storage}
\begin{tabular}{|c|c|ccccc|c|c|}

\hline
\multirow{2}{*}{\textbf{\begin{tabular}[c]{@{}c@{}}Number of\\  Samples\end{tabular}}} & \multirow{2}{*}{\textbf{\begin{tabular}[c]{@{}c@{}}Model\end{tabular}}} & \multicolumn{5}{c|}{\textbf{Time Required (s) for Training the models}}                                                                                               & \multirow{2}{*}{\textbf{\begin{tabular}[c]{@{}c@{}}Inference\\ Time (s)\end{tabular}}} & \multicolumn{1}{l|}{\multirow{2}{*}{\textbf{\begin{tabular}[c]{@{}l@{}}Memory\\ Required\end{tabular}}}} \\ \cline{3-7}
                                                                                       &                                                                                      & \multicolumn{1}{c|}{\textbf{$1^{st}$ Fold}} & \multicolumn{1}{c|}{\textbf{$2^{nd}$ Fold}} & \multicolumn{1}{c|}{\textbf{$3^{rd}$ Fold}} & \multicolumn{1}{c|}{\textbf{$4^{th}$ Fold}} & \textbf{$5^{th}$ Fold} &                                                                                               & \multicolumn{1}{l|}{}                                                                                                  \\ \hline
                                                                                       & \textbf{DNN}                                                                         & \multicolumn{1}{c|}{11.215}          & \multicolumn{1}{c|}{10.994}          & \multicolumn{1}{c|}{13.418}          & \multicolumn{1}{c|}{21.693}          & 12.223          & 0.26417                                                                                       & 40 KB                                                                                                                  \\
                                                                                       & \textbf{PINN}                                                                        & \multicolumn{1}{c|}{12.970}          & \multicolumn{1}{c|}{22.370}          & \multicolumn{1}{c|}{17.891}          & \multicolumn{1}{c|}{18.063}          & 16.883          & 0.33131                                                                                       & 40 KB                                                                                                                  \\
\textbf{4000}                                                                          & \textbf{StackDNN}                                                                    & \multicolumn{1}{c|}{118.229}         & \multicolumn{1}{c|}{97.498}          & \multicolumn{1}{c|}{98.634}          & \multicolumn{1}{c|}{101.169}         & 100.175         & 0.26417                                                                                       & 360 KB                                                                                                                 \\
                                                                                       & \textbf{StackPINN}                                                                   & \multicolumn{1}{c|}{100.464}         & \multicolumn{1}{c|}{100.062}         & \multicolumn{1}{c|}{101.656}         & \multicolumn{1}{c|}{101.386}         & 102.123         & 0.33131                                                                                       & 360 KB                                                                                                                 \\
                                                                                       & \textbf{XGBoost}                                                                         & \multicolumn{1}{c|}{11.250}          & \multicolumn{1}{c|}{11.791}          & \multicolumn{1}{c|}{41.222}          & \multicolumn{1}{c|}{31.340}          & 61.009          & 0.27845                                                                                       & 3900 KB                                                                                                                \\ \hline
                                                                                       & \textbf{DNN}                                                                         & \multicolumn{1}{c|}{22.230}          & \multicolumn{1}{c|}{21.886}          & \multicolumn{1}{c|}{22.479}          & \multicolumn{1}{c|}{22.337}          & 21.955          & 0.26594                                                                                       & 40 KB                                                                                                                  \\
                                                                                       & \textbf{PINN}                                                                        & \multicolumn{1}{c|}{25.407}          & \multicolumn{1}{c|}{25.273}          & \multicolumn{1}{c|}{25.082}          & \multicolumn{1}{c|}{25.214}          & 25.093          & 0.32936                                                                                       & 40 KB                                                                                                                  \\
\textbf{8235}                                                                          & \textbf{StackDNN}                                                                    & \multicolumn{1}{c|}{200.484}         & \multicolumn{1}{c|}{200.760}         & \multicolumn{1}{c|}{201.941}         & \multicolumn{1}{c|}{201.911}         & 200.889         & 0.26594                                                                                       & 360 KB                                                                                                                 \\
                                                                                       & \textbf{StackPINN}                                                                   & \multicolumn{1}{c|}{204.296}         & \multicolumn{1}{c|}{203.725}         & \multicolumn{1}{c|}{203.401}         & \multicolumn{1}{c|}{204.594}         & 205.888         & 0.32936                                                                                       & 360 KB                                                                                                                 \\
                                                                                       & \textbf{XGBoost}                                                                         & \multicolumn{1}{c|}{7.963}           & \multicolumn{1}{c|}{8.183}           & \multicolumn{1}{c|}{8.119}           & \multicolumn{1}{c|}{8.323}           & 8.295           & 0.28983                                                                                       & 4232 KB                                                                                                                \\ \hline
\end{tabular}
\end{table}

\begin{table}[ht]
\begin{minipage}{.5\linewidth}
\centering
\caption{Median error for the test dataset}
\label{tab:Median_error}
\scalebox{0.8}{
\begin{tabular}{|c|l|r|r|r|r|}
\hline
\textbf{}              & \multicolumn{1}{c|}{\textbf{\begin{tabular}[c]{@{}c@{}}Model\end{tabular}}} & \multicolumn{1}{c|}{\textbf{$\Delta r$ (km)}} & \multicolumn{1}{c|}{\textbf{$\Delta v$ (km/s)}} & \multicolumn{1}{c|}{\textbf{$\Delta m$ (kg)}} & \multicolumn{1}{c|}{\textbf{$\Delta \epsilon$}} \\ \hline
\textbf{}              & \textbf{DNN}                                                                             & 0.674506                                           & 2.027031                                               & 0.394279                                      & 1.076511                                        \\
\textbf{}              & \textbf{PINN}                                                                            & 0.473144                                           & 2.010821                                               & 0.178281                                      & 0.767108                                        \\
\textbf{4000}          & \textbf{StackDNN}                                                                        & 0.152939                                           & 1.764392                                               & -0.017718                                     & 0.034151                                        \\
\textbf{}              & \textbf{StackPINN}                                                                       & 0.158544                                           & 2.013816                                               & 0.002398                                      & 0.030450                                        \\
\textbf{}              & \textbf{XGBoost}                                                                         & 0.189282                                           & 0.617731                                               & 0.000088                                      & -0.035985                                       \\ \hline
\textbf{}              & \textbf{DNN}                                                                             & 0.344972                                           & 1.873850                                               & 0.272428                                      & 0.059095                                        \\
\textbf{}              & \textbf{PINN}                                                                            & 0.165065                                           & 1.816507                                               & 0.019991                                      & 0.036363                                        \\
\textbf{8235}          & \textbf{StackDNN}                                                                        & 0.154050                                           & 2.475177                                               & -0.000288                                     & 0.033286                                        \\
\textbf{}              & \textbf{StackPINN}                                                                       & 0.156674                                           & 3.232480                                               & -0.000230                                     & 0.045742                                        \\
\textbf{}              & \textbf{XGBoost}                                                                         & 0.185646                                           & 0.825226                                               & 0.000249                                      & -0.040400                                       \\ \hline
\multicolumn{1}{|l|}{} & \textbf{Lagrange}                                                                        & 0.155496                                           & 1.558974                                               & 0.490096                                      & 0.290463                                        \\ \hline
\end{tabular}
}
\end{minipage}
\hfill
\begin{minipage}{.5\linewidth}
\centering
\caption{Interquartile range of the errors for the test dataset}
\label{tab:iqr_table}
\scalebox{0.8}{
\begin{tabular}{|c|l|r|r|r|r|}
\hline
\textbf{}              & \multicolumn{1}{c|}{\textbf{\begin{tabular}[c]{@{}c@{}}Model\end{tabular}}} & \multicolumn{1}{c|}{\textbf{$\Delta r$ (km)}} & \multicolumn{1}{c|}{\textbf{$\Delta v
$ (km/s)}} & \multicolumn{1}{c|}{\textbf{$\Delta m$ (kg)}} & \multicolumn{1}{c|}{\textbf{$\Delta \epsilon$}} \\ \hline
\textbf{}              & \textbf{DNN}                                                                             & 0.186429                                           & 2.010423                                               & 0.002568                                      & 0.231941                                        \\
\textbf{}              & \textbf{PINN}                                                                            & 0.163924                                           & 2.238280                                               & 0.001413                                      & 0.201985                                        \\
\textbf{4000}          & \textbf{StackDNN}                                                                        & 0.073994                                           & 2.002598                                               & 0.001530                                      & 0.185718                                        \\
\textbf{}              & \textbf{StackPINN}                                                                       & 0.069802                                           & 2.313196                                               & 0.002019                                      & 0.184570                                        \\
\textbf{}              & \textbf{XGBoost}                                                                         & 0.126994                                           & 0.562895                                               & 0.001296                                      & 0.103200                                        \\ \hline
\textbf{}              & \textbf{DNN}                                                                             & 0.205209                                           & 1.814919                                               & 0.213681                                      & 0.188687                                        \\
\textbf{}              & \textbf{PINN}                                                                            & 0.088477                                           & 1.894849                                               & 0.001394                                      & 0.183433                                        \\
\textbf{8235}          & \textbf{StackDNN}                                                                        & 0.067543                                           & 1.750979                                               & 0.001412                                      & 0.178207                                        \\
\textbf{}              & \textbf{StackPINN}                                                                       & 0.085123                                           & 1.482897                                               & 0.001205                                      & 0.181471                                        \\
\textbf{}              & \textbf{XGBoost}                                                                         & 0.091459                                           & 0.840224                                               & 0.001372                                      & 0.127520                                        \\ \hline
\multicolumn{1}{|l|}{} & \textbf{Lagrange}                                                                        & 0.073416                                           & 2.193088                                               & 0.001430                                      & 0.183433                                        \\ \hline
\end{tabular}
}
\end{minipage}
\end{table}

Table \ref{tab:Median_error} and Table \ref{tab:iqr_table} summarize the inference of the selected trained models along with the Lagrange method for the LEMUR collision dataset. From these tables, we can conclude that the PINN with 2 hidden layers provides better performance to resource requirement trade-off for tracking untracked space debris from a non-destructive and inelastic collision with an active satellite than the DNN, Stacked neural networks, and classical methods. We have found that the XGBoost method provides better results than the DNN and the PINN. However, this method requires approximately 100 times more memory than the DNN and PINN-based techniques. As a result, the PINN-based model would be more suitable for on-board deployment on the
active spacecraft. Additionally, the XGBoost is a black box model, whereas the PINN is intrinsically explainable due to the
physics-based loss function.




\section*{Conclusion}
We have formulated a problem of space debris position, velocity, mass, and coefficient of restitution estimation from a non-destructive inelastic collision event under the assumption that the occurrence of the collision event is known. We have shown that this problem can be posed as a function approximation problem; hence, an ML model can approximate the function. We proposed a loss function based on Newton's law of gravitation to design a PINN to solve this problem. We have also presented an algorithm for selecting the velocity of space debris for a collision simulation. Using simulated collision data, we have trained the DNN, PINN, StackDNN, and StackPINN with varied numbers of hidden layers and the XGBoost model. We have also compared the estimation performance of these ML-based methods with the Lagrange multiplier-based optimization technique. 

The results demonstrate that the PINN with 2 hidden layers performs better estimation than the DNN with 2 hidden layers and the classical method. Additionally, the DNN performance becomes similar to the PINN when the number of hidden layers increases and when Stacked neural network architecture is used. However, this increase in the model complexity does not improve the estimation performance compared to the PINN with 2 hidden layers. It is also observed that the XGBoost models provide better estimation performance than the neural network-based counterparts. It is noteworthy to mention that the XGBoost model is a black-box model, and as a consequence, the output of the model is not explainable. On the other hand, the PINN, being trained using a loss function derived from the laws of physics, provides explainability of the outputs. In addition, the XGBoost model requires significantly higher memory than the proposed PINN. However, it is anticipated that model-agnostic explanation techniques, for example, SHapley Additive exPlanations (SHAP) \cite{lundberg2017unified} can be utilised to explain the XGBoost model in this context and can be deployed on the ground or onboard spacecraft leveraging the state-of-the-art embedded systems. A physics-informed approach for the explainability of the gradient boosting technique \cite{fang2023ensemble} can also be explored for tracking space debris by observing a collision.

The problem presented in the article can be further extended to the estimation of space debris position, velocity, mass, and coefficient of restitution using a sequence of tracking data of the active satellite. The Recurrent Neural Network (RNN) or transformer may be suitable for this case. However, it will require the generation of a large number of sequences of tracking data for each collision to train such models. RNNs, Transformers, and other such models tend to have significantly more inference times, model sizes, and complexity. Further research is necessary to study the performance of these models in the problem under consideration. It is anticipated that the application of the PINN can be further extended to trajectory extraction of the debris after a break-up event. In addition, the velocity sampling formulation can be used to derive a velocity probability density function for a given position in space, which can be used to design new filters for selecting candidate space debris for collision assessment as well as enhancing the collision probability computation.


\begin{thebibliography}{10}
\urlstyle{rm}
\expandafter\ifx\csname url\endcsname\relax
  \def\url#1{\texttt{#1}}\fi
\expandafter\ifx\csname urlprefix\endcsname\relax\def\urlprefix{URL }\fi
\expandafter\ifx\csname doiprefix\endcsname\relax\def\doiprefix{DOI: }\fi
\providecommand{\bibinfo}[2]{#2}
\providecommand{\eprint}[2][]{\url{#2}}

\bibitem{montaruliAdaptiveTrackEstimation2022}
\bibinfo{author}{Montaruli, M.~F.} \emph{et~al.}
\newblock \bibinfo{journal}{\bibinfo{title}{Adaptive track estimation on a radar array system for space surveillance}}.
\newblock {\emph{\JournalTitle{Acta Astronautica}}} \textbf{\bibinfo{volume}{198}}, \bibinfo{pages}{111--123}, \doiprefix\url{10.1016/j.actaastro.2022.05.051} (\bibinfo{year}{2022}).

\bibitem{tanAnalysisIridium332013}
\bibinfo{author}{Tan, A.}, \bibinfo{author}{Zhang, T.~X.} \& \bibinfo{author}{Dokhanian, M.}
\newblock \bibinfo{journal}{\bibinfo{title}{Analysis of the {{Iridium}} 33 and {{Cosmos}} 2251 {{Collision}} using {{Velocity Perturbations}} of the {{Fragments}}}}.
\newblock {\emph{\JournalTitle{Advances in Aerospace Science and Applications}}} \textbf{\bibinfo{volume}{3}} (\bibinfo{year}{2013}).

\bibitem{dattaOpedDamageCanadarm22021}
\bibinfo{author}{Datta, A.}
\newblock \bibinfo{title}{Op-ed | {{Damage}} to {{Canadarm2}} on {{ISS}} once again highlights space debris problem} (\bibinfo{year}{2021}).

\bibitem{kelsoWhatHappenedBLITS2013}
\bibinfo{author}{Kelso, T.~S.} \emph{et~al.}
\newblock \bibinfo{journal}{\bibinfo{title}{What {{Happened}} to {{BLITS}}? {{An Analysis}} of the 2013 {{Jan}} 22 {{Event}}}}.
\newblock {\emph{\JournalTitle{Proceedings of the Advanced Maui Optical and Space Surveillance Technologies Conference}}} \textbf{\bibinfo{volume}{4}} (\bibinfo{year}{2013}).

\bibitem{braun2020drama}
\bibinfo{author}{Braun, V.}, \bibinfo{author}{Funke, Q.}, \bibinfo{author}{Lemmens, S.} \& \bibinfo{author}{Sanvido, S.}
\newblock \bibinfo{journal}{\bibinfo{title}{Drama 3.0-upgrade of esa’s debris risk assessment and mitigation analysis tool suite}}.
\newblock {\emph{\JournalTitle{Journal of Space Safety Engineering}}} \textbf{\bibinfo{volume}{7}}, \bibinfo{pages}{206--212} (\bibinfo{year}{2020}).

\bibitem{braun2021recent}
\bibinfo{author}{Braun, V.}, \bibinfo{author}{Horstmann, A.}, \bibinfo{author}{Lemmens, S.}, \bibinfo{author}{Wiedemann, C.} \& \bibinfo{author}{B{\"o}ttcher, L.}
\newblock \bibinfo{title}{Recent developments in space debris environment modelling, verification and validation with master}.
\newblock In \emph{\bibinfo{booktitle}{8th European Conference on Space Debris}} (\bibinfo{organization}{ESA Space Debris Office Darmstadt, Germany}, \bibinfo{year}{2021}).

\bibitem{lopez-calleComparisonCubesatMicrosat2023}
\bibinfo{author}{{Lopez-Calle}, I.} \& \bibinfo{author}{Franco, A.~I.}
\newblock \bibinfo{journal}{\bibinfo{title}{Comparison of cubesat and microsat catastrophic failures in function of radiation and debris impact risk}}.
\newblock {\emph{\JournalTitle{Scientific Reports}}} \textbf{\bibinfo{volume}{13}}, \bibinfo{pages}{385}, \doiprefix\url{10.1038/s41598-022-27327-z} (\bibinfo{year}{2023}).

\bibitem{cellettiReconnectingGroupsSpace2021}
\bibinfo{author}{Celletti, A.}, \bibinfo{author}{Pucacco, G.} \& \bibinfo{author}{Vartolomei, T.}
\newblock \bibinfo{journal}{\bibinfo{title}{Reconnecting groups of space debris to their parent body through proper elements}}.
\newblock {\emph{\JournalTitle{Scientific Reports}}} \textbf{\bibinfo{volume}{11}}, \bibinfo{pages}{22676}, \doiprefix\url{10.1038/s41598-021-02010-x} (\bibinfo{year}{2021}).

\bibitem{mDeepNeuralNetworkbased2023}
\bibinfo{author}{M, H.}, \bibinfo{author}{Dave, A.~A.}, \bibinfo{author}{Singh, G.}, \bibinfo{author}{Buduru, A.~B.} \& \bibinfo{author}{Biswas, S.~K.}
\newblock \bibinfo{title}{A {{Deep Neural Network-based Space}} debris trajectory prediction after an elastic collision event}.
\newblock In \emph{\bibinfo{booktitle}{{{SMOPS Conference}} 2023}} (\bibinfo{address}{{Bangalore, India}}, \bibinfo{year}{2023}).

\bibitem{song_deep_2022}
\bibinfo{author}{Song, J.}, \bibinfo{author}{Rondao, D.} \& \bibinfo{author}{Aouf, N.}
\newblock \bibinfo{journal}{\bibinfo{title}{Deep learning-based spacecraft relative navigation methods: {A} survey}}.
\newblock {\emph{\JournalTitle{Acta Astronautica}}} \textbf{\bibinfo{volume}{191}}, \bibinfo{pages}{22--40}, \doiprefix\url{10.1016/j.actaastro.2021.10.025} (\bibinfo{year}{2022}).

\bibitem{becktor2022robust}
\bibinfo{author}{Becktor, J.} \emph{et~al.}
\newblock \bibinfo{title}{Robust vision-based multi-spacecraft guidance navigation and control using cnn-based pose estimation}.
\newblock In \emph{\bibinfo{booktitle}{2022 IEEE Aerospace Conference (AERO)}}, \bibinfo{pages}{1--10} (\bibinfo{organization}{IEEE}, \bibinfo{year}{2022}).

\bibitem{petit2012vision}
\bibinfo{author}{Petit, A.}, \bibinfo{author}{Marchand, E.} \& \bibinfo{author}{Kanani, K.}
\newblock \bibinfo{title}{Vision-based detection and tracking for space navigation in a rendezvous context}.
\newblock In \emph{\bibinfo{booktitle}{Int. Symp. on Artificial Intelligence, Robotics and Automation in Space, i-SAIRAS}} (\bibinfo{year}{2012}).

\bibitem{kaluthantrigeCNNbasedImageProcessing2023}
\bibinfo{author}{Kaluthantrige, A.}, \bibinfo{author}{Feng, J.} \& \bibinfo{author}{{Gil-Fern{\'a}ndez}, J.}
\newblock \bibinfo{journal}{\bibinfo{title}{{{CNN-based Image Processing}} algorithm for autonomous optical navigation of {{Hera}} mission to the binary asteroid {{Didymos}}}}.
\newblock {\emph{\JournalTitle{Acta Astronautica}}} \textbf{\bibinfo{volume}{211}}, \bibinfo{pages}{60--75}, \doiprefix\url{10.1016/j.actaastro.2023.05.029} (\bibinfo{year}{2023}).

\bibitem{stevensonPredictingEffectsKinetic2022}
\bibinfo{author}{Stevenson, E.}, \bibinfo{author}{Martinez, R.}, \bibinfo{author}{{Rodriguez-Fernandez}, V.} \& \bibinfo{author}{Camacho, D.}
\newblock \bibinfo{title}{Predicting the effects of kinetic impactors on asteroid deflection using end-to-end deep learning}.
\newblock In \emph{\bibinfo{booktitle}{2022 {{IEEE Congress}} on {{Evolutionary Computation}} ({{CEC}})}}, \bibinfo{pages}{1--8}, \doiprefix\url{10.1109/CEC55065.2022.9870215} (\bibinfo{year}{2022}).

\bibitem{stevensonBenchmarkingDeepLearning2023}
\bibinfo{author}{Stevenson, E.}, \bibinfo{author}{{Rodriguez-Fernandez}, V.}, \bibinfo{author}{Urrutxua, H.} \& \bibinfo{author}{Camacho, D.}
\newblock \bibinfo{journal}{\bibinfo{title}{Benchmarking deep learning approaches for all-vs-all conjunction screening}}.
\newblock {\emph{\JournalTitle{Advances in Space Research}}} \textbf{\bibinfo{volume}{72}}, \bibinfo{pages}{2660--2675}, \doiprefix\url{10.1016/j.asr.2023.01.036} (\bibinfo{year}{2023}).

\bibitem{sanchezIntelligentDecisionSupport2023}
\bibinfo{author}{S{\'a}nchez, L.} \& \bibinfo{author}{Vasile, M.}
\newblock \bibinfo{journal}{\bibinfo{title}{Intelligent decision support for collision avoidance manoeuvre planning under uncertainty}}.
\newblock {\emph{\JournalTitle{Advances in Space Research}}} \textbf{\bibinfo{volume}{72}}, \bibinfo{pages}{2627--2648}, \doiprefix\url{10.1016/j.asr.2022.09.023} (\bibinfo{year}{2023}).

\bibitem{Raissi_2018}
\bibinfo{author}{Raissi, M.} \& \bibinfo{author}{Karniadakis, G.~E.}
\newblock \bibinfo{journal}{\bibinfo{title}{Hidden physics models: Machine learning of nonlinear partial differential equations}}.
\newblock {\emph{\JournalTitle{Journal of Computational Physics}}} \textbf{\bibinfo{volume}{357}}, \bibinfo{pages}{125--141}, \doiprefix\url{10.1016/j.jcp.2017.11.039} (\bibinfo{year}{2018}).

\bibitem{raissiPhysicsinformedNeuralNetworks2019}
\bibinfo{author}{Raissi, M.}, \bibinfo{author}{Perdikaris, P.} \& \bibinfo{author}{Karniadakis, G.~E.}
\newblock \bibinfo{journal}{\bibinfo{title}{Physics-informed neural networks: {{A}} deep learning framework for solving forward and inverse problems involving nonlinear partial differential equations}}.
\newblock {\emph{\JournalTitle{Journal of Computational Physics}}} \textbf{\bibinfo{volume}{378}}, \bibinfo{pages}{686--707}, \doiprefix\url{10.1016/j.jcp.2018.10.045} (\bibinfo{year}{2019}).

\bibitem{karniadakis2021physics}
\bibinfo{author}{Karniadakis, G.~E.} \emph{et~al.}
\newblock \bibinfo{journal}{\bibinfo{title}{Physics-informed machine learning}}.
\newblock {\emph{\JournalTitle{Nature Reviews Physics}}} \textbf{\bibinfo{volume}{3}}, \bibinfo{pages}{422--440} (\bibinfo{year}{2021}).

\bibitem{goswami2022physicsinformed}
\bibinfo{author}{Goswami, S.}, \bibinfo{author}{Bora, A.}, \bibinfo{author}{Yu, Y.} \& \bibinfo{author}{Karniadakis, G.~E.}
\newblock \bibinfo{title}{Physics-informed deep neural operator networks} (\bibinfo{year}{2022}).
\newblock \eprint{2207.05748}.

\bibitem{brakeComprehensiveSetImpact2017a}
\bibinfo{author}{Brake, M. R.~W.}, \bibinfo{author}{Reu, P.~L.} \& \bibinfo{author}{Aragon, D.~S.}
\newblock \bibinfo{journal}{\bibinfo{title}{A {{Comprehensive Set}} of {{Impact Data}} for {{Common Aerospace Metals}}}}.
\newblock {\emph{\JournalTitle{Journal of Computational and Nonlinear Dynamics}}} \textbf{\bibinfo{volume}{12}}, \doiprefix\url{10.1115/1.4036760} (\bibinfo{year}{2017}).

\bibitem{schwagerCoefficientRestitutionLinear2007}
\bibinfo{author}{Schwager, T.} \& \bibinfo{author}{Poeschel, T.}
\newblock \bibinfo{title}{Coefficient of restitution and linear dashpot model revisited} (\bibinfo{year}{2007}).
\newblock \eprint{cond-mat/0701278}.

\bibitem{vallado2001fundamentals}
\bibinfo{author}{Vallado, D.~A.}
\newblock \emph{\bibinfo{title}{Fundamentals of astrodynamics and applications}}, vol.~\bibinfo{volume}{12} (\bibinfo{publisher}{Springer Science \& Business Media}, \bibinfo{year}{2001}).

\bibitem{theastropycollaborationAstropyProjectSustaining2022}
\bibinfo{author}{Price-Whelan, A.~M.} \emph{et~al.}
\newblock \bibinfo{title}{The {{Astropy Project}}: {{Sustaining}} and {{Growing}} a {{Community-oriented Open-source Project}} and the {{Latest Major Release}} (v5.0) of the {{Core Package}}}, \doiprefix\url{10.3847/1538-4357/ac7c74} (\bibinfo{year}{2022}).
\newblock \eprint{2206.14220}.

\bibitem{rodriguezPoliastroAstrodynamicsLibrary2016}
\bibinfo{author}{Rodr{\'i}guez, J. L.~C.}, \bibinfo{author}{Eichhorn, H.} \& \bibinfo{author}{McLean, F.}
\newblock \bibinfo{title}{Poliastro: An astrodynamics library written in python with fortran performance}.
\newblock In \emph{\bibinfo{booktitle}{6th {{International Conference}} on {{Astrodynamics Tools}} and {{Techniques}}}} (\bibinfo{year}{2016}).

\bibitem{chen2016xgboost}
\bibinfo{author}{Chen, T.} \& \bibinfo{author}{Guestrin, C.}
\newblock \bibinfo{title}{Xgboost: A scalable tree boosting system}.
\newblock In \emph{\bibinfo{booktitle}{Proceedings of the 22nd acm sigkdd international conference on knowledge discovery and data mining}}, \bibinfo{pages}{785--794} (\bibinfo{year}{2016}).

\bibitem{mhaskar2016deep}
\bibinfo{author}{Mhaskar, H.~N.} \& \bibinfo{author}{Poggio, T.}
\newblock \bibinfo{journal}{\bibinfo{title}{Deep vs. shallow networks: An approximation theory perspective}}.
\newblock {\emph{\JournalTitle{Analysis and Applications}}} \textbf{\bibinfo{volume}{14}}, \bibinfo{pages}{829--848} (\bibinfo{year}{2016}).

\bibitem{lundberg2017unified}
\bibinfo{author}{Lundberg, S.~M.} \& \bibinfo{author}{Lee, S.-I.}
\newblock \bibinfo{journal}{\bibinfo{title}{A unified approach to interpreting model predictions}}.
\newblock {\emph{\JournalTitle{Advances in neural information processing systems}}} \textbf{\bibinfo{volume}{30}} (\bibinfo{year}{2017}).

\bibitem{fang2023ensemble}
\bibinfo{author}{Fang, Z.}, \bibinfo{author}{Wang, S.} \& \bibinfo{author}{Perdikaris, P.}
\newblock \bibinfo{journal}{\bibinfo{title}{Ensemble learning for physics informed neural networks: a gradient boosting approach}}.
\newblock {\emph{\JournalTitle{arXiv preprint arXiv:2302.13143}}}  (\bibinfo{year}{2023}).

\end{thebibliography}

\section*{Acknowledgements}

We acknowledge the National Supercomputing Mission (NSM) for providing computing resources of ‘PARAM PRAVEGA’ at SERC Building IISc Main Campus Bangalore, which is implemented by C-DAC and supported by the Ministry of Electronics and Information Technology (MeitY) and Department of Science and Technology (DST), Government of India. We also acknowledge the support of the Infosys Foundation for conducting our research. The authors would like to acknowledge the support of U R Rao Satellite Centre, Indian Space Research Organisation (ISRO).

\section*{Author contributions statement}
S.K.B. conceived the research idea, S.K.B., A.B.B., and V.K. contributed to developing the methodology, H.M., G.S. and S.K.B. conducted the simulation experiments and analysed results. All authors reviewed the manuscript.
\section*{Funding}
This research is supported by National Super Computing Mission  (NSM) - HPC Applications, Grant no.
DST/NSM/R\&D\_HPC\_Applications/2021/03.17.
\section*{Additional information}
\textbf{Competing interests}\\
The authors declare no competing interests.\\
\newline
\newline
\textbf{Data Availability}\\
The datasets generated and analysed during the current study are available in the \textbf{collisionAI} repository, \\
\href{https://github.com/HarshaSSL/collisionAI}{https://github.com/HarshaSSL/collisionAI}
\newline
\newline
\textbf{Correspondence} and requests for materials should be addressed to H.M.

\section*{Appendix}
\label{sec:appendix}

\subsection*{Models trained using 4000 samples}
\begin{figure}[ht]
    \centering
    \includegraphics[width=\textwidth]{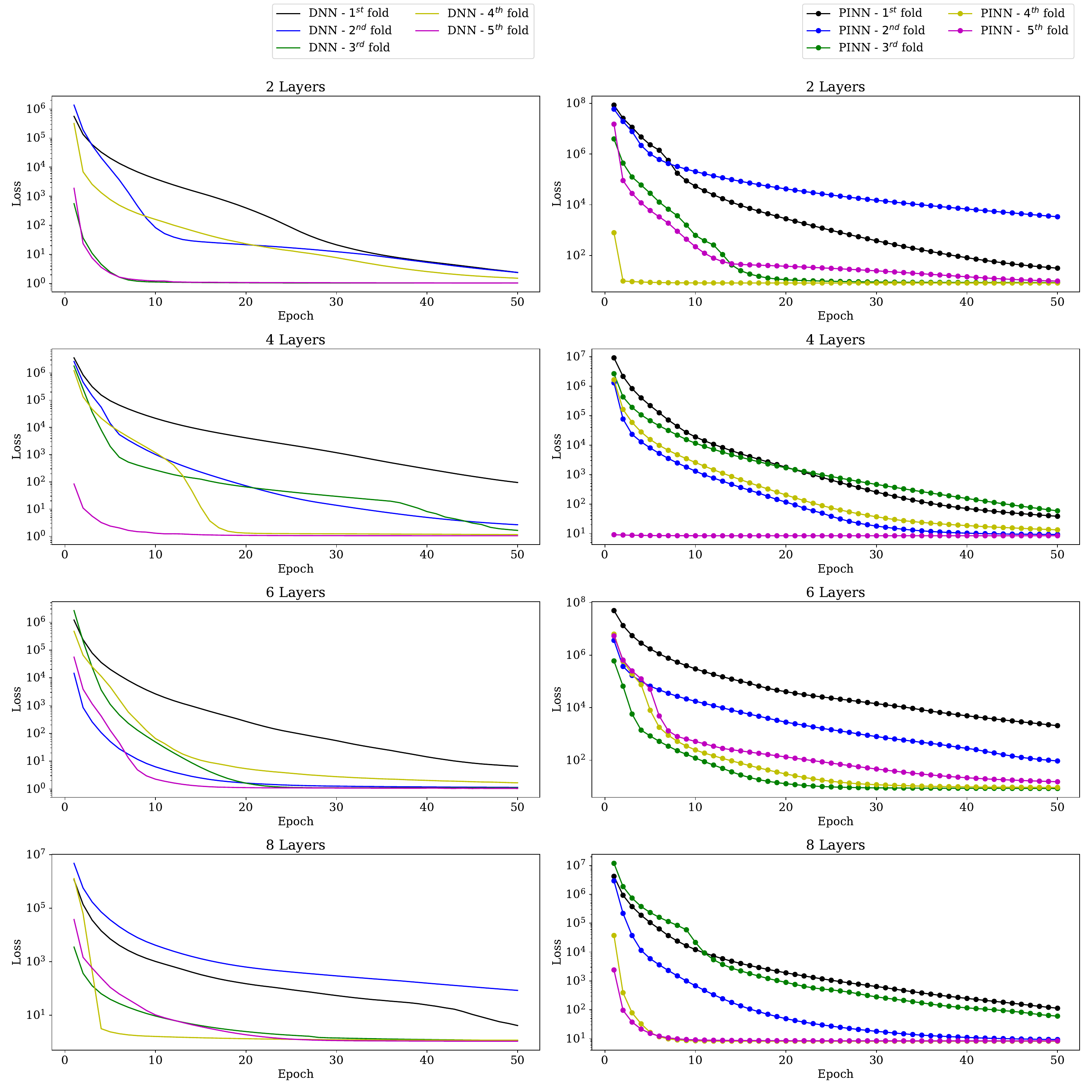}
    \caption{Training loss for DNN and PINN models in 5-Fold Cross-validation using 4000 samples and 50 epochs}
    \label{fig:dnn_pinn_loss_plot(4000)}
\end{figure}

\begin{figure}[ht]
    \centering
    \includegraphics[width=\textwidth]{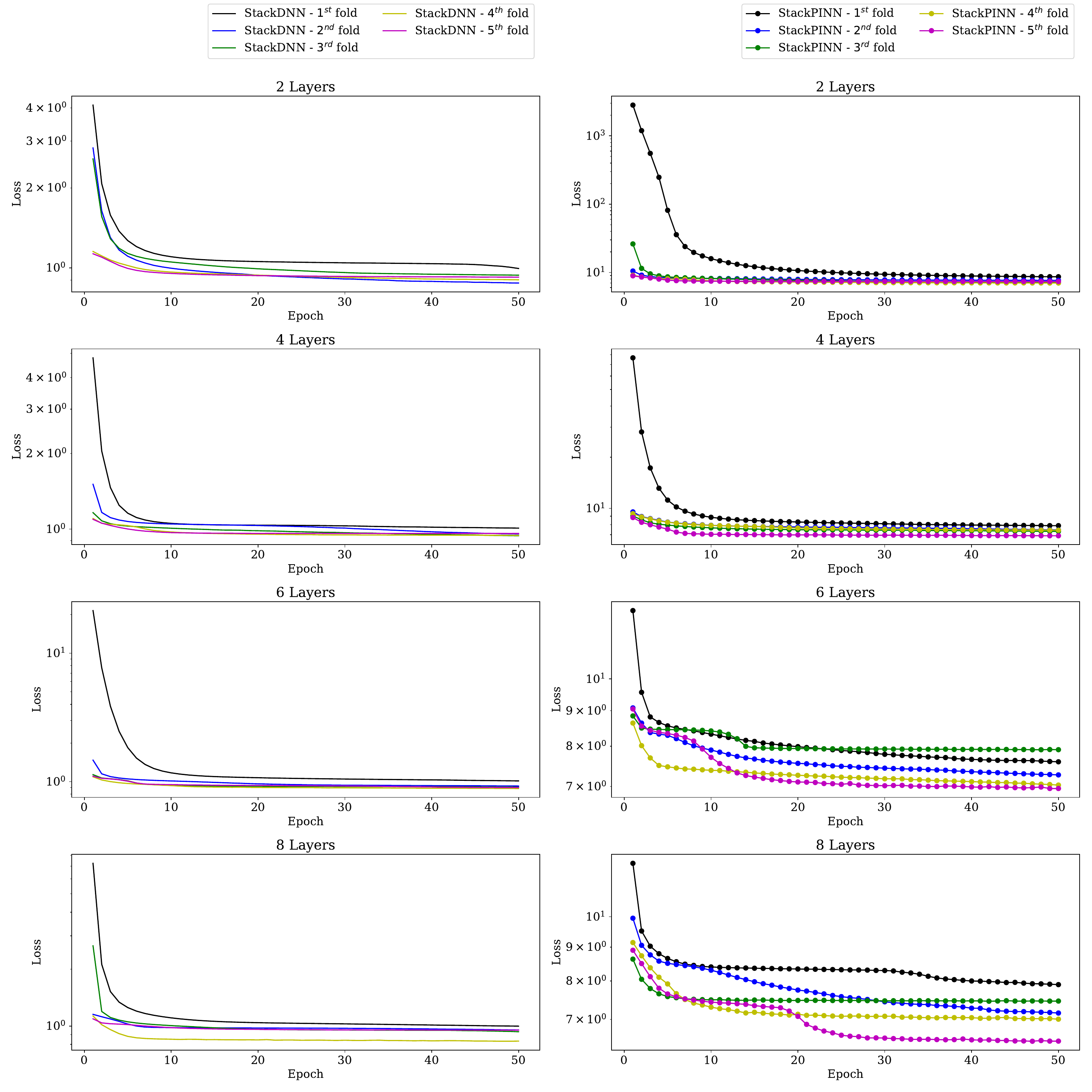}
    \caption{Training loss for StackDNN and StackPINN models in 5-Fold Cross-validation using 4000 samples and 50 epochs}
    \label{fig:stackdnn_stackpinn_loss_plot(4000)}
\end{figure}

\begin{figure}[ht]
    \centering
    \includegraphics[width=\textwidth]{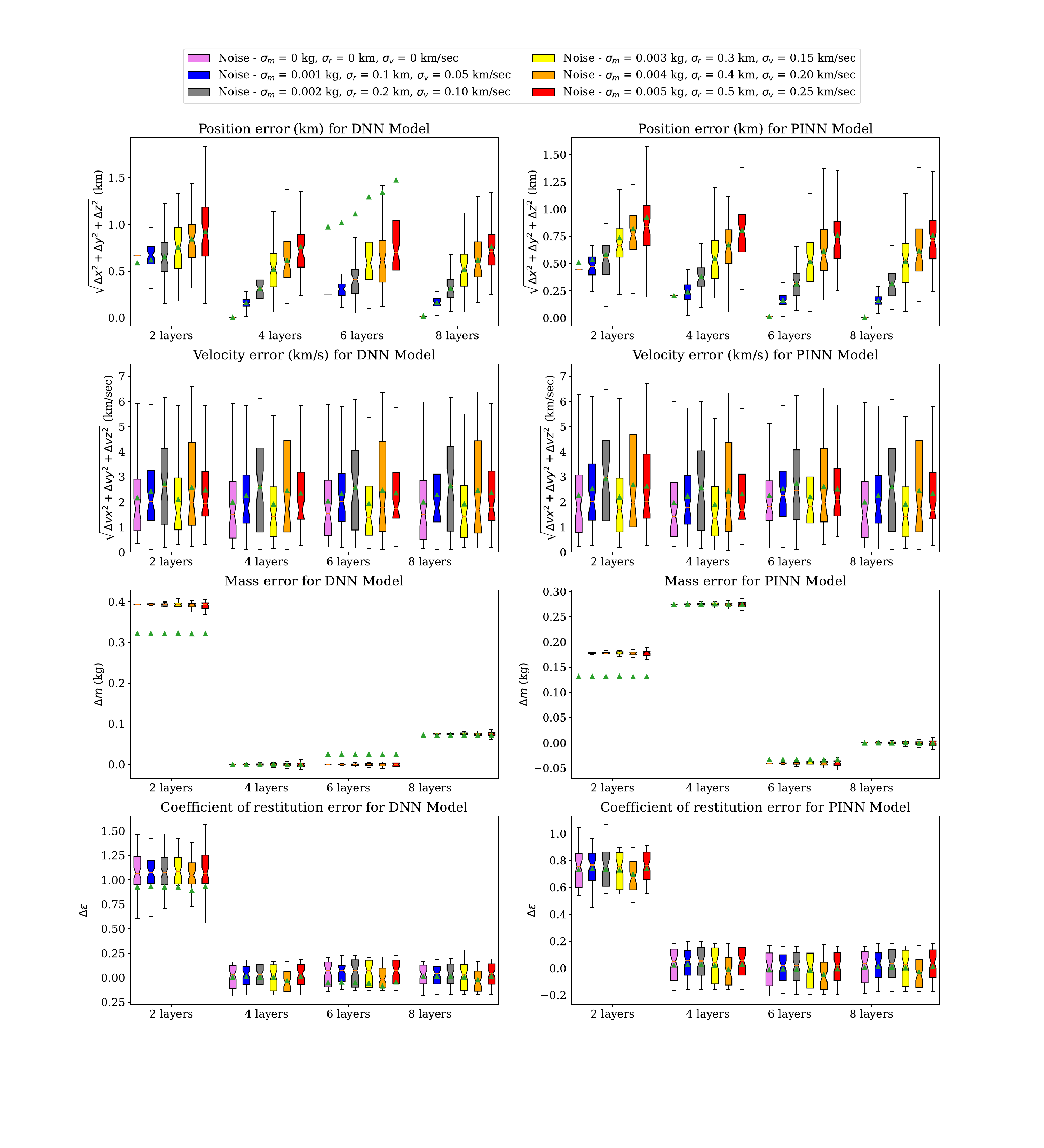}
    \caption{Error comparison of DNN and PINN models trained using 4000 samples and 50 epochs for multiple noise variations in LEMUR Satellites inelastic collision data}
    \label{fig:StackDNN(full)_error_vs_noise(4000)}
\end{figure}

\begin{figure}[ht]
    \centering
    \includegraphics[width=\textwidth]{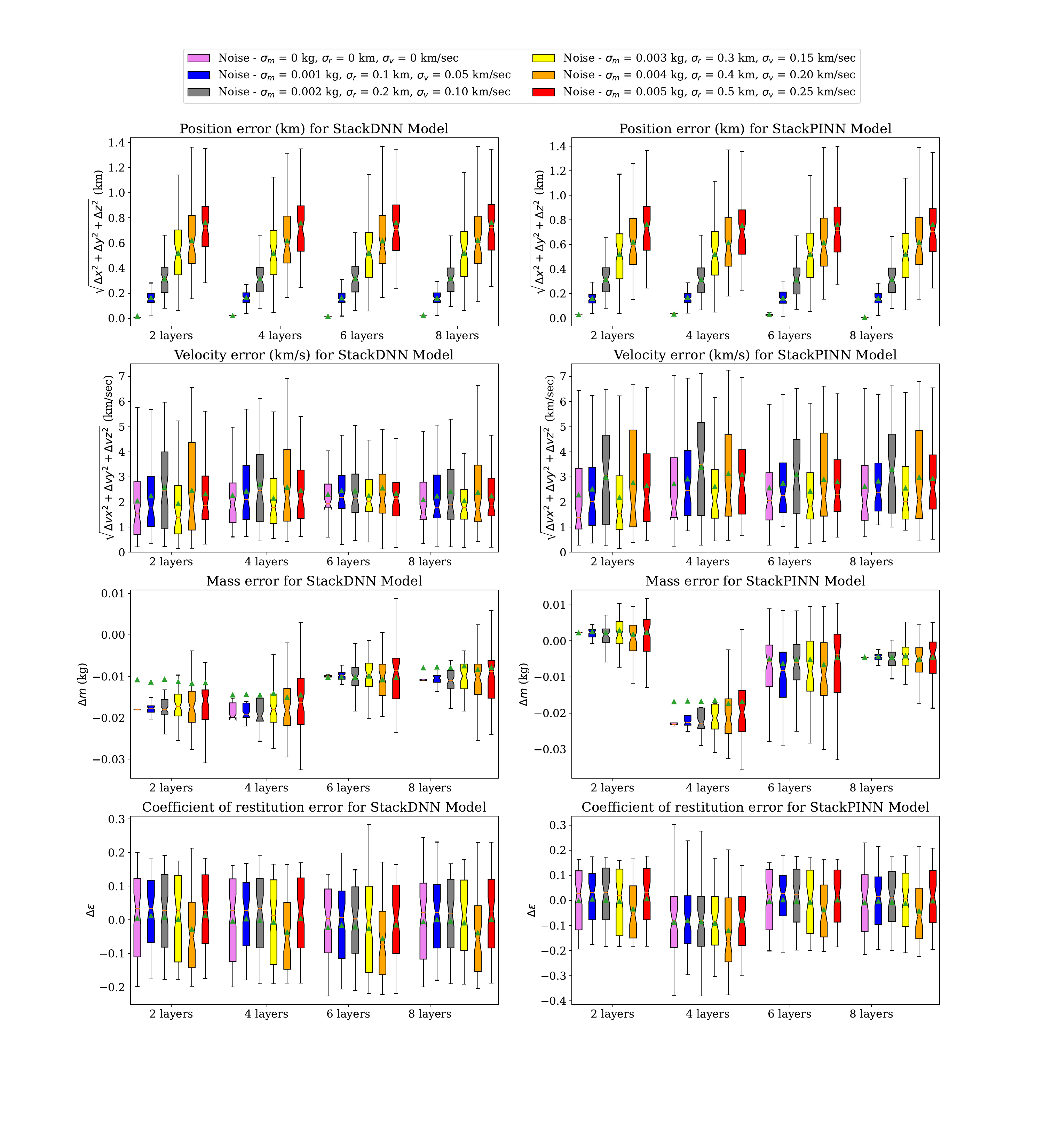}
    \caption{Error comparison of StackDNN and StackPINN models trained using 4000 samples and 50 epochs for multiple noise variations in LEMUR satellites inelastic collision data}
    \label{fig:StackPINN(full)_error_vs_noise(4000)}
\end{figure}

\end{document}